\documentclass[a4paper,fleqn,usenatbib]{mnras}

\usepackage{newtxtext,newtxmath}

\usepackage[T1]{fontenc}
\usepackage{ae,aecompl}

\pdfoutput=1

\usepackage{epsfig,graphicx}	
\usepackage{amsmath}	
\usepackage{amssymb}	
\usepackage{natbib}
\usepackage{bm}
\usepackage{comment}
\usepackage{dcolumn}
\usepackage{subfig}

\title[Compact Jet Variability]
{Radio Frequency Timing Analysis of the Compact Jet in the Black Hole X-ray Binary Cygnus X-1}

\author[A.J. Tetarenko et al.]{A.J. Tetarenko$^{1,2}$\thanks{E-mail: a.tetarenko@eaobservatory.org},
P. Casella$^3$,
J.C.A. Miller-Jones$^4$,
G.R. Sivakoff$^1$,
\newauthor
B.E. Tetarenko$^{1,5}$,
T.J. Maccarone$^6$,
P. Gandhi$^7$,
and S. Eikenberry$^{8}$
\\
$^{1}$Department of Physics, University of Alberta, CCIS 4-181, Edmonton, AB T6G 2E1, Canada\\
$^2$East Asian Observatory, 660 N. A'ohoku Place, University
Park, Hilo, Hawaii 96720, USA\\
$^{3}$INAF-Osservatorio Astronomico di Roma, Via Frascati 33, I-00078 Monteporzio Catone, Italy\\
$^{4}$International Centre for Radio Astronomy Research -- Curtin University, GPO Box U1987, Perth, WA 6845, Australia\\
$^5$Department of Astronomy, University of Michigan, 1085 South University Avenue, Ann Arbor, MI 48109, USA\\
$^{6}$Department of Physics and Astronomy, Texas Tech University, Lubbock, Texas 79409-1051, USA\\
$^{7}$School of Physics and Astronomy, University of Southampton, Southampton, SO17 1BJ, UK\\
$^{8}$Department of Astronomy, University of Florida, 211 Bryant Space Science Center, Gainesville, FL 32611, USA\\
}

\date{Accepted XXX. Received YYY; in original form ZZZ}

\pubyear{2016}

\begin{document}
\label{firstpage}
\pagerange{\pageref{firstpage}--\pageref{lastpage}}
\maketitle

\begin{abstract}
We present simultaneous multi-band radio and X-ray observations of the black hole X-ray binary Cygnus X-1, taken with the Karl G. Jansky Very Large Array and the Nuclear Spectroscopic Telescope Array. With these data, we detect clear flux variability consistent with emission from a variable compact jet. To probe how the variability signal propagates down the jet flow, we perform detailed timing analyses of our data. We find that the radio jet emission shows no significant power at Fourier frequencies $f\gtrsim0.03$ Hz (below $\sim30$ sec timescales), and that the higher frequency radio bands (9/11 GHz) are strongly correlated over a range of timescales, displaying a roughly constant time lag with Fourier frequency of a few tens of seconds. However, in the lower frequency radio bands (2.5/3.5 GHz) we find a significant loss of coherence over the same range of timescales. Further, we detect a correlation between the X-ray/radio emission, measuring time lags between the X-ray/radio bands on the order of tens of minutes. We use these lags to solve for the compact jet speed, finding that the Cyg X-1 jet is more relativistic than usually assumed for compact jets, where $\beta=0.92^{+0.03}_{-0.06}$, ($\Gamma=2.59^{+0.79}_{-0.61}$). Lastly, we constrain how the jet size scale changes with frequency, finding a shallower relation ($\propto \nu^{-0.4}$) than predicted by simple jet models ($\propto \nu^{-1}$), and estimate a jet opening angle of $\phi\sim0.4-1.8$ degrees. With this study we have developed observational techniques designed to overcome the challenges of radio timing analyses and created the tools needed to connect rapid radio jet variability properties to internal jet physics.

\end{abstract}


\section{Introduction}
Black holes drive the most powerful outflows in the Universe, from the kiloparsec-scale jets launched by the most massive black holes in Active Galactic Nuclei (AGN), to the smaller-scale jets launched by their stellar-mass analogues,  black hole X-ray binaries (BHXBs). 
BHXBs are often the targets of jet studies, as these accreting binary systems rapidly evolve through bright outburst phases (typically lasting days to months), providing a real time view of jet behaviour.
At the beginning of an outburst, the BHXB is typically in a hard accretion state, where there exists a compact, relativistic
jet, an inner radiatively inefficient accretion flow \citep{naryi95,mar05}, and a slightly truncated accretion disc \citep{mr06, fenbelgal04}. The compact jets emit most prominently at radio, sub-mm and infrared frequencies \citep{fen01,cor02aa,chat,rus06,tetarenkoa2015} as a result of partially self-absorbed synchrotron emission \citep{blandford79}, where higher frequencies probe regions closer to the black hole. 
On the other hand, the accretion flow (in the hard state) primarily emits in the X-rays, and is dominated by thermal comptonization of photons from the accretion disc (or synchrotron self-compton) in the inner accretion flow/jet base region (although the irradiated outer accretion disc can also contribute, and in some cases dominate, the emission at infrared, optical, and UV frequencies; \citealt{vanp1994,vanp96,d07}).
This X-ray emission is known to be strongly variable (fractional rms 10--50\%), on as short as sub-second timescales (power at Fourier frequencies as high as hundreds of Hz; \citealt{vanderklis}). 

While broad-band spectral measurements and high resolution radio imaging studies with Very Long Baseline Interferometry (VLBI; e.g. \citealt{stir01}) are traditionally used to constrain jet properties (e.g., speed and geometry), and probe the connection between the jet and accretion flows (e.g., \citealt{van13,rus14}), time domain observations offer a promising new way to address the key open questions in jet research \citep{utt15}.
Detecting and characterizing rapid flux variability in jet emission from multiple BHXBs can allow us to probe detailed jet properties that are difficult, if not impossible, to measure by other means. For instance, measuring the shortest timescale over which the compact jet flux is significantly changing provides a direct measure of the jet size scale at different observing frequencies. At its best, high resolution VLBI can only image the jets from BHXBs (which are located at kiloparsec distances) down to AU size scales. In fact, in the few published cases where a compact jet was resolved in the axial direction (GRS 1915+105 and Cygnus X-1;  \citealt{stir01,dhaw00}), VLBI has failed to resolve structure perpendicular to the jet axis, suggesting jet widths on the order of sub-AU scales. Therefore, while determining the
cross-section and opening angle of BHXB jets is beyond current imaging capabilities, detecting jet variability on second timescales, probing scales as small as mAU (for emitting regions travelling at light speed), could be used to recover new information on jet geometry.

Additionally, as the jet flow propagates away from the black hole, optical depth effects cause lower frequency emission to appear as a delayed version of high frequency emission \citep{blandford79,hj88,falcke95}. Through measuring the time delay between low level variability features at different frequencies, we can estimate the compact jet speed, for
which there are currently no direct measurements \citep{utt15}. 
Further, detecting correlated variability across a wide range of frequencies, probing both the jet and accretion flow emission, can allow us to effectively track accreting matter from inflow to outflow, revealing how variations in the accretion flow manifest themselves further downstream in the jet. Recent work has shown evidence of correlations between optical/infrared and X-ray variability on sub-second timescales \citep{mal03,cas10,kalam16,gan17,vinc18,mal18}, suggesting that variations
in the accretion flow could subsequently drive variability in the jet emission. 
Directly linking changes in the accretion flow with changes in jet emission at different scales (where different frequencies probe different distances along the jet axis) provides unparalleled insight into the {sequence of events leading to jet launching and acceleration}.

Broad-band variability measurements also allow for detailed tests of both standard and new jet theory models. In particular, \cite{blandford79} predict that the jet size scale is inversely proportional to observing frequency ({$z\propto \nu^{-1}$, where $z$ represents the distance from where the jet is launched to the $\tau=1$ surface}).
Additionally, recent models
predict that jet variability is driven by the injection of discrete shells of plasma at the base of the jet with variable speeds \citep{tur04,jam10,mal14,drap15,drap17,mal18}. In these works, the behaviour of these shells (traced by the jet variability properties) is directly linked to the X-ray power density spectrum (quantifying the amplitude of X-ray variability at different timescales).
As the timescale of the jet variability depends on the shock speed and shell thickness in this model,
detecting correlated variability over a wide frequency range could disentangle these parameters.

To date, compact jet emission in BHXBs at radio frequencies has been observed to vary over a range of timescales (minutes to months). 
While the longer timescale variations have been tracked and well characterized in many systems, this is not the case for the short ($<1$ hour) timescale variations, for which {there are only a handful of sources with detections (e.g., variations detected on minute to hour timescales; \citealt{poolf97,mir98,fenpoo00a,corb0,fenray00,millerjtur09,curr14,tetarenkoa2015})}. Further, little effort has been made to analyze this short timescale radio variability (e.g., \citealt{nip05} present the only radio frequency Fourier domain study, where the shortest timescales probed were days), {to search for variability at higher sub-mm frequencies probing the base of the jet close to the black hole (and bridging the gap between the radio and OIR frequencies)}, or to connect radio variability properties (e.g., amplitudes, timescales) with internal jet physics.

While time-resolved observations are a staple for BHXB studies at shorter wavelengths, there are
many challenges that accompany such studies at long wavelengths (radio, sub-mm).
In particular, it can be difficult to disentangle intrinsic source variations from atmospheric or telescope gain variations (the radio sky at $>10$ GHz is sparse enough that in-beam comparison sources will be rare), observations often involve routinely cycling between observing a target source and a calibrator, and in most cases only one frequency band can be sampled at a time. 
These obstacles prevent continuous observations of the jet and introduce artificial variability signals in the data, both of which complicate time-domain analyses. Further, until recently, most telescopes were not sensitive enough, nor capable of taking the rapid data to probe second timescales.
However, with today's more sensitive interferometric arrays, which also offer observing modes that allow for the use of sub-arrays\footnote{We note that \citealt{mir98} were the first to utilize the sub-arraying technique to simultaneously observe an XB (GRS 1915+105) at multiple radio frequencies with the VLA.}, sub-second time resolution, and custom non-periodic target$+$calibrator scan design, we can lift the limitations presented above and accurately sample BHXB jets at radio frequencies in the time domain.
Here we present new simultaneous, high time resolution, multi-band radio and X-ray observations of the compact jet in the BHXB Cygnus X-1, and perform a detailed study of rapid (probing second to hour timescales) compact jet variability at radio frequencies.

 \subsection{Cygnus X-1}
Cygnus X-1 (hereafter Cyg X-1) is a high-mass X-ray binary (containing an O-type super-giant companion; \citealt{gies86}) discovered in the X-rays by the UHURU satellite in 1971 \citep{tan72}.
It is located at a distance of $1.86\pm0.12$ kpc \citep{re11}, with an orbital period of 5.6 d \citep{holt79} and an inclination angle of $27.1\pm0.8$ \citep{or11} degrees to our line of sight. {For this work we will assume that the jet axis is perpendicular to the accretion disc. This is a valid assumption for Cyg X-1, as the low measured proper motion suggests that this system did not receive a strong kick during the black hole formation process \citep{mirb03,re11}. However, X-ray reflection modelling (e.g., \citealt{park2015,tom2018}) suggests that the black hole spin axis (and hence the jet axis) may be misaligned with the orbital plane ($\sim40$ degrees vs. $\sim27$ degrees). Therefore, the inclination of the jet axis could be closer to $40$ degrees.} {We also note that the distance cited above is a radio parallax distance, and there is a discrepancy between this radio parallax distance and the {\em Gaia} DR2 \citep{gaiadr2} distance of $2.38^{+0.20}_{-0.17}$ kpc (where the derived {\em Gaia} distance value is dependent on the prior distribution used; see \citealt{gand18} for details). We use the radio parallax distance throughout this paper, and we find that the choice of distance value (radio parallax or {\em Gaia}) does not significantly affect our conclusions (see \S\ref{sec:speed}).}

Cyg X-1 spends the majority of its time in a canonical hard accretion state (e.g., \citealt{ling83,mi92,wilm06,zdz2,grin13}), where the radio through sub-mm spectrum is very flat ($f_\nu\propto\nu^\alpha$, where $\alpha\simeq0$), displaying an average flux density of $\sim 15$ mJy \citep{fen2000}. However, Cyg X-1 has shown small-scale radio frequency variability, with amplitudes  up to 20--30\% of the average flux density, on timescales of hours to months. The shorter timescale variability (hours--days) is thought to be linked to changes in the mass accretion rate, while the longer timescale variability (months) has been attributed to jet precession (although astrometric fits to VLBI data of Cyg X-1 may rule out long term precession), and orbital modulation originating from variable absorption by the stellar wind of the companion star  \citep{poo99cyg,bro02,gle04,nip05,pan06,wil07}.
The hard state radio jet in Cyg X-1 has been resolved along the jet axis out to scales of $\sim$15 mas at 8.4 GHz \citep{stir01}, but has not been resolved in the direction perpendicular to the jet flow, constraining the jet opening angle to be $<2$ degrees.

Cyg X-1 is also known to be strongly variable in the X-rays during the hard state ({fractional rms 40--50\% over the $\sim$0.002-128 Hz range}), showing power over a large range of Fourier frequencies from mHz to over 100 Hz \citep{pott3}.
Despite the extended time Cyg X-1 spends in the hard state, the source has been known to make occasional transitions to a softer accretion state, and was in such a soft state (with no compact jet) from 2010 until late 2015 \citep{grinberg2011,grinberg14,grin15}.

In 2016 February, while Cyg X-1 was in a well-established hard accretion state, we obtained simultaneous, high time resolution, multi-band radio and X-ray observations with NSF's Karl G. Jansky Very Large Array (VLA) and NASA's Nuclear Spectroscopic Telescope Array ({\em NuSTAR}).
In \S\ref{sec:data} we describe the data collection and reduction processes. In \S\ref{sec:results} we present high time resolution light curves, cross-correlation functions, and Fourier domain analyses of the radio and X-ray emission we detect from Cyg X-1. In \S\ref{sec:discuss} we discuss the time domain properties of the jet emission from Cyg X-1 and place constraints on jet speed, geometry, and size scales. A summary of the results is presented in \S\ref{sec:sum}.

\section{Observations and data analysis}
\label{sec:data}

\begin{figure*}
   \centering
   \includegraphics[width=1.3\columnwidth]{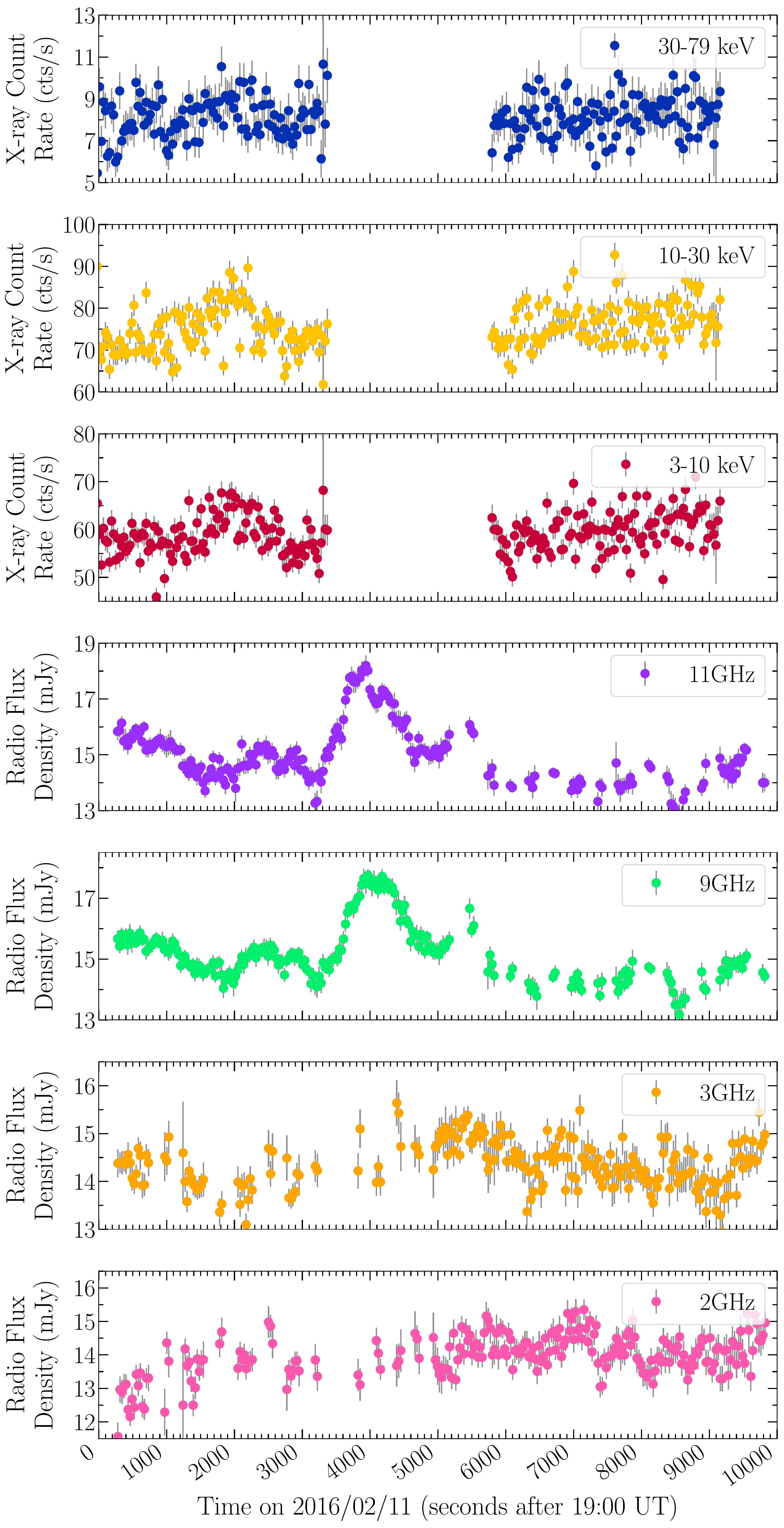}
 \caption{ \label{fig:cyglc}Multi-frequency light curves of the BHXB Cyg X-1 (with 30 sec time-bins). {The panels from top to bottom show light curves for progressively decreasing frequency/energy bands; the top three panels display {\em NuSTAR} X-ray light curves in three different energy bands and the bottom four panels display VLA radio light curves in four different frequency bands.} The {\em NuSTAR} light curves displayed here are a {subset} of the full light curve, shown in Figure~\ref{fig:Xlc} in Appendix~\ref{sec:xraylc_ex}. We observe flux variability in both the radio and X-ray light curves, in the form of small-scale flaring activity. The radio variability appears to be of lower amplitude and more smoothed out in the lower frequency bands.
 }
 \end{figure*}

\subsection{VLA radio observations}
\label{sec:radio_red}
Cyg X-1 was observed with the VLA (Project Code: 16A-241) on 2016 February 11, for a total on-source observation time of 2.7 hours. The array was in the C configuration at the time of our observations, where we split the full array into 2 sub-arrays of 13 and 14 antennas. Observations in each sub-array were made with the 8-bit samplers in S ($2-4\,{\rm GHz}$) and X ($8-12\,{\rm GHz}$) band. Each band was comprised of 2 base-bands, with 8 spectral windows of 64 2-MHz channels each, giving a total bandwidth of 1.024 GHz per base-band. We chose to use the $2-4\,{\rm GHz}$ and $8-12\,{\rm GHz}$ bands as together they allow us to avoid the difficulties of observing at higher frequencies, and act as a compromise between high frequencies that probe closer to the black hole and lower frequencies that are more sensitive with the VLA. Further, the $8-12\,{\rm GHz}$ band provides a pair of widely separated base-bands. The sub-array setup allows us to push to shorter correlator dump times than would be possible if we were using the full array. In these observations, we set a 0.25-second correlator dump time, providing the highest time resolution possible, while staying within the standard $25\, {\rm MB\, s}^{-1}$ data rate limit. {For comparison, our sub-array setup allows us to record data fast enough to probe timescales down to hundreds of milli-seconds, while real-time commensal transient surveys such as V-FASTR \citep{wayth11} or REALFAST \citep{law18}  can identify variable sources on much faster, milli-second timescales}.
Further, we implemented a custom non-periodic target$+$calibrator cycle for each sub-array. 
This custom cycle began with 30 minutes for the dummy setup scan and standard calibration in two bands. Following this standard calibration, each sub-array alternated observing Cyg X-1 and calibrators, such that we obtained uninterrupted data of Cyg X-1 in the $8-12\,{\rm GHz}$ band for the first 75 minutes of the observations, and uninterrupted data of Cyg X-1 in the $2-4\,{\rm GHz}$ band for the second 75 minutes of the observations (with sparser coverage of the second band across each 75-min period).
We hand-set target$+$calibrator scans to between 9 and 15 minutes (with approximately logarithmic separation). 
Flagging, calibration and imaging of the data were carried out within the Common Astronomy Software Application package ({\sc casa} v5.1.2; \citealt{mc07}) using standard procedures outlined in the \textsc{casa} Guides\footnote{\url{https://casaguides.nrao.edu}} for VLA data reduction (i.e., a priori flagging, setting the flux density scale, initial phase calibration, solving for antenna-based delays, bandpass calibration, gain calibration, scaling the amplitude gains, and final target flagging). We used 3C48 (J0137$+$331) as a flux/bandpass calibrator and J2015$+$3710 as a phase calibrator. To obtain high time resolution flux density measurements, we used our custom {\sc casa} variability measurement scripts\footnote{These scripts are publicably available on github; \url{https://github.com/Astroua/AstroCompute\_Scripts}}, where flux densities of the source in each time bin were measured by fitting a point source in the image plane (with the \texttt{imfit} task). When imaging each time bin we used a natural weighting scheme to maximize sensitivity and did not perform any self-calibration. Note that we only analyze radio light curves on timescales as short as 1 sec in this work (despite the 0.25 sec time resolution), as we found no significant power on sub-second timescales (see \S\ref{sec:psd}). Additionally, to check that any variability observed in these Cyg X-1 radio frequency light curves is dominated by intrinsic variations in the source, and not due to atmospheric or instrumental effects, we also ran our calibrator sources through these scripts (see Appendix~\ref{sec:appendix} for details).

\subsection{NuSTAR X-ray observations}
Cyg X-1 was observed with {\em NuSTAR} \citep{harrison2013} on 2016 Febuary 11, for a total exposure time of 13.5 ks (ObsID 90101020002). The {\em NuSTAR} telescope consists of two co-aligned focal plane modules; FPMA and FPMB.
The data were reduced using the {\em NuSTAR} data analysis software ({\sc nustardas} v1.8.0) within the {\sc heasoft} software package (v6.22\footnote{\url{http://heasarc.nasa.gov/lheasoft/}}) following standard procedures\footnote{{\em NuSTAR} data analysis procedures are detailed at \url{https://heasarc.gsfc.nasa.gov/docs/nustar/analysis/}}. We first used the \texttt{nupipeline} task to filter the observations for passages through the South Atlantic Anomaly and produce cleaned event files. Then we extracted light curves from both FPMA and FPMB using a circular region with a $60$ arcsec radius centred on the source. Similarly, background light curves were extracted from a 100 arcsec radius circular region centred on a source-free region on each detector. Lastly, the {\sc heasoft} task \texttt{lcmath} was used to create final background subtracted light curves.
Note that we extracted light curves in the 3--10 keV, 10--30 keV, 30--79 keV, and full 3--79 keV energy bands, on time scales as short as 1 sec (matching our radio observations; see \S\ref{sec:radio_red}), for the analysis presented in this paper.

\section{Results}
\label{sec:results}
\subsection{Light curves}

 \begin{figure*}
   \centering
   \subfloat{\includegraphics[width=0.95\columnwidth]{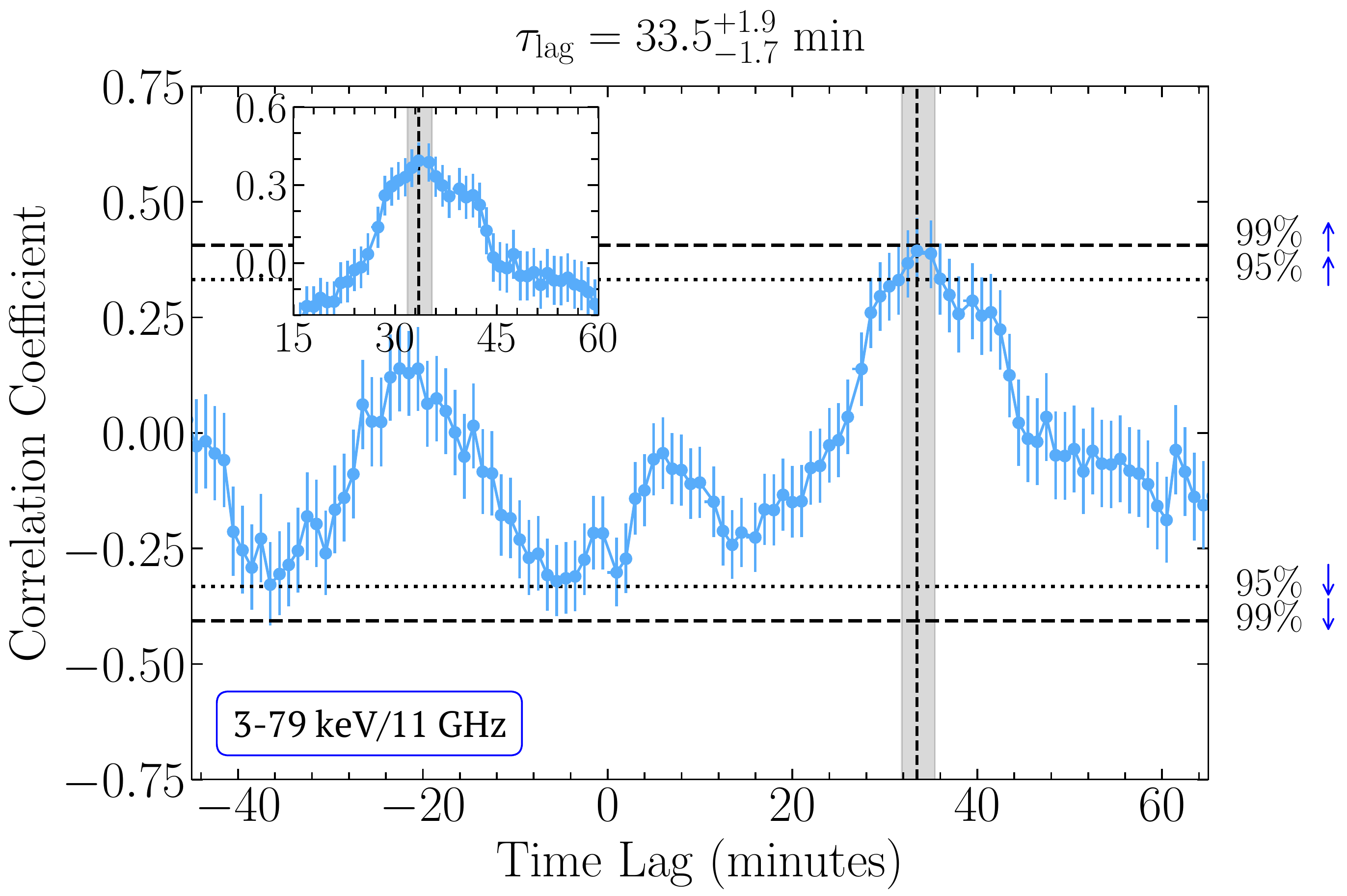}}\quad\quad
    \subfloat{\includegraphics[width=0.95\columnwidth]{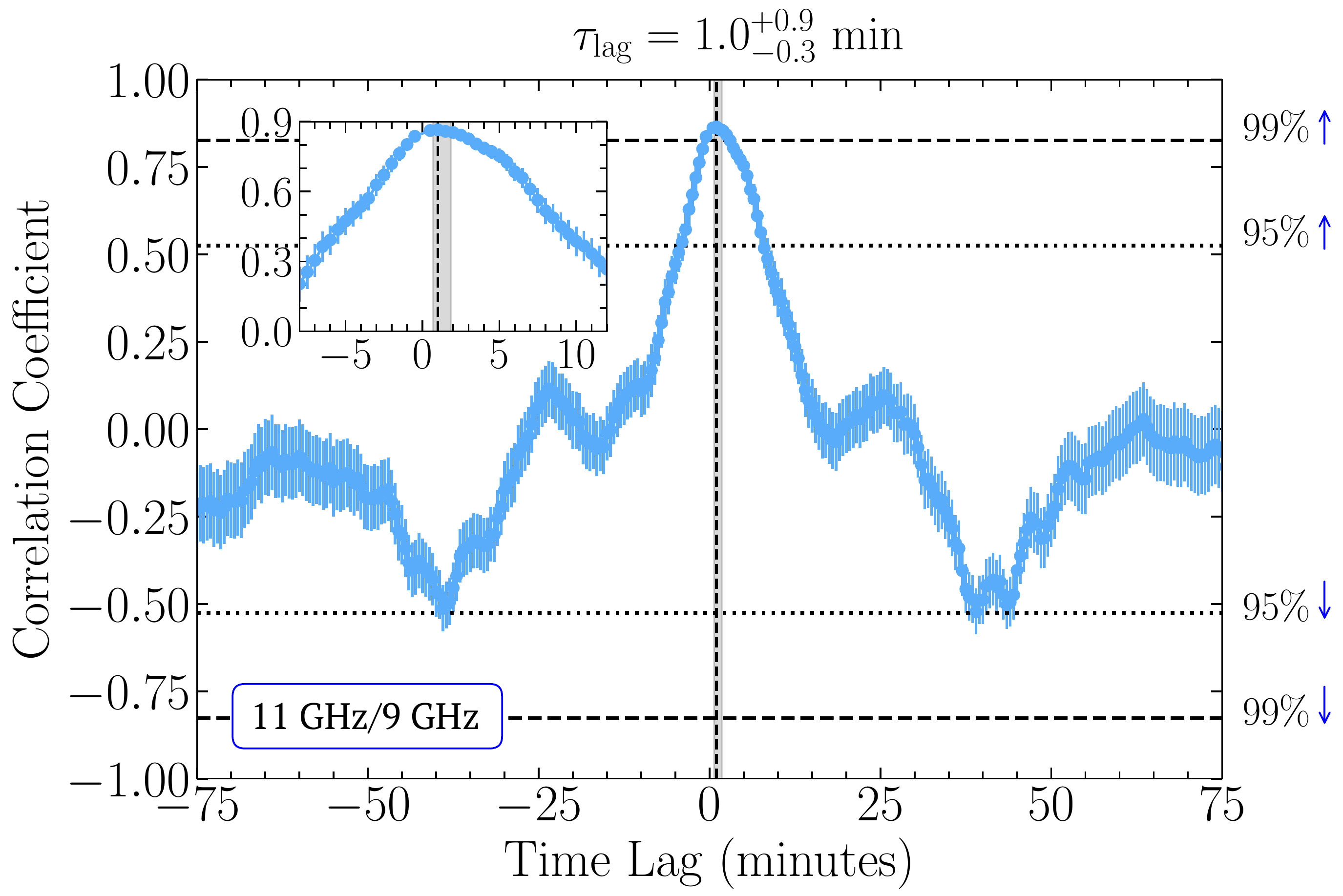}}\\
   \subfloat{ \includegraphics[width=0.95\columnwidth]{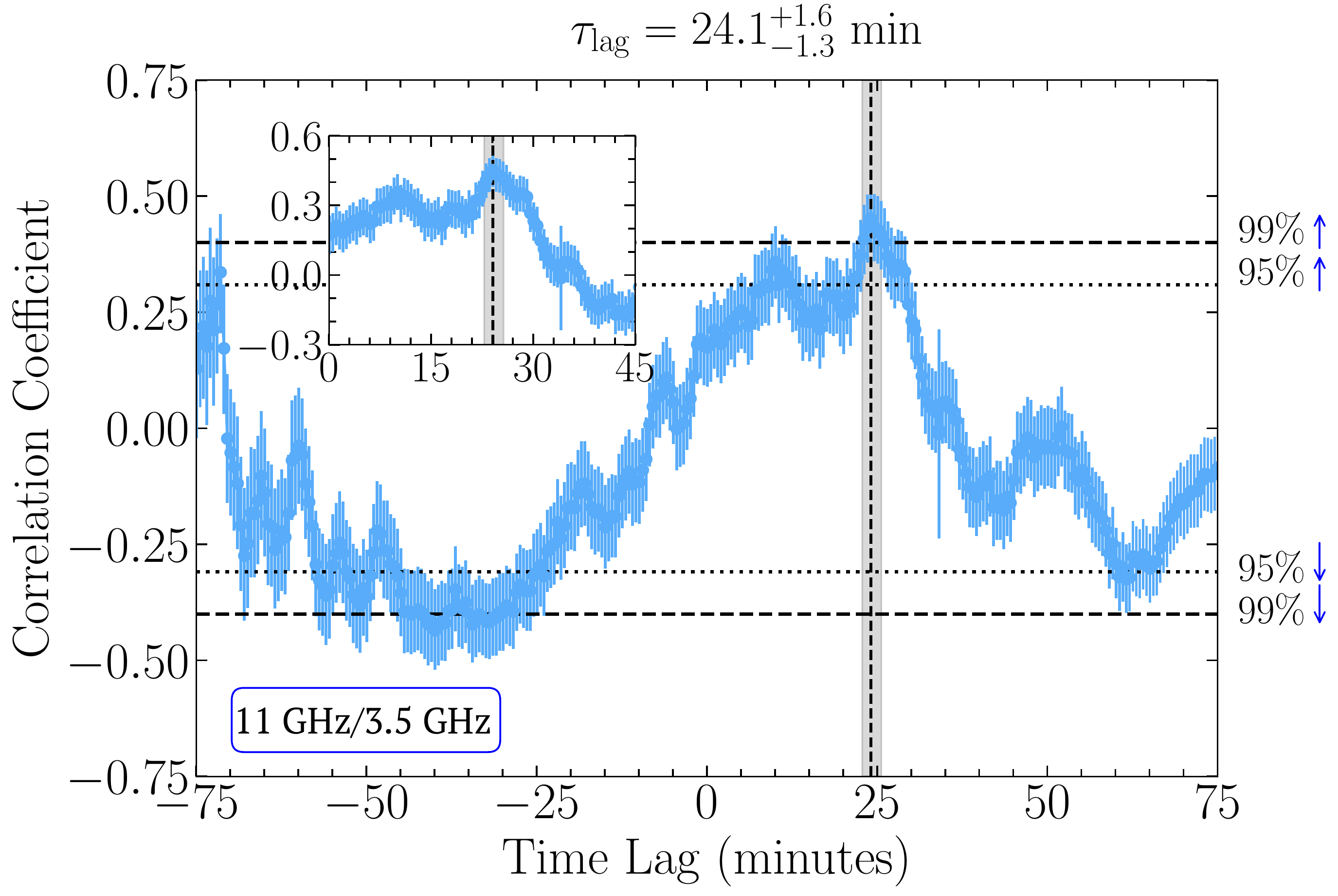}} \quad\quad
   \subfloat{ \includegraphics[width=0.95\columnwidth]{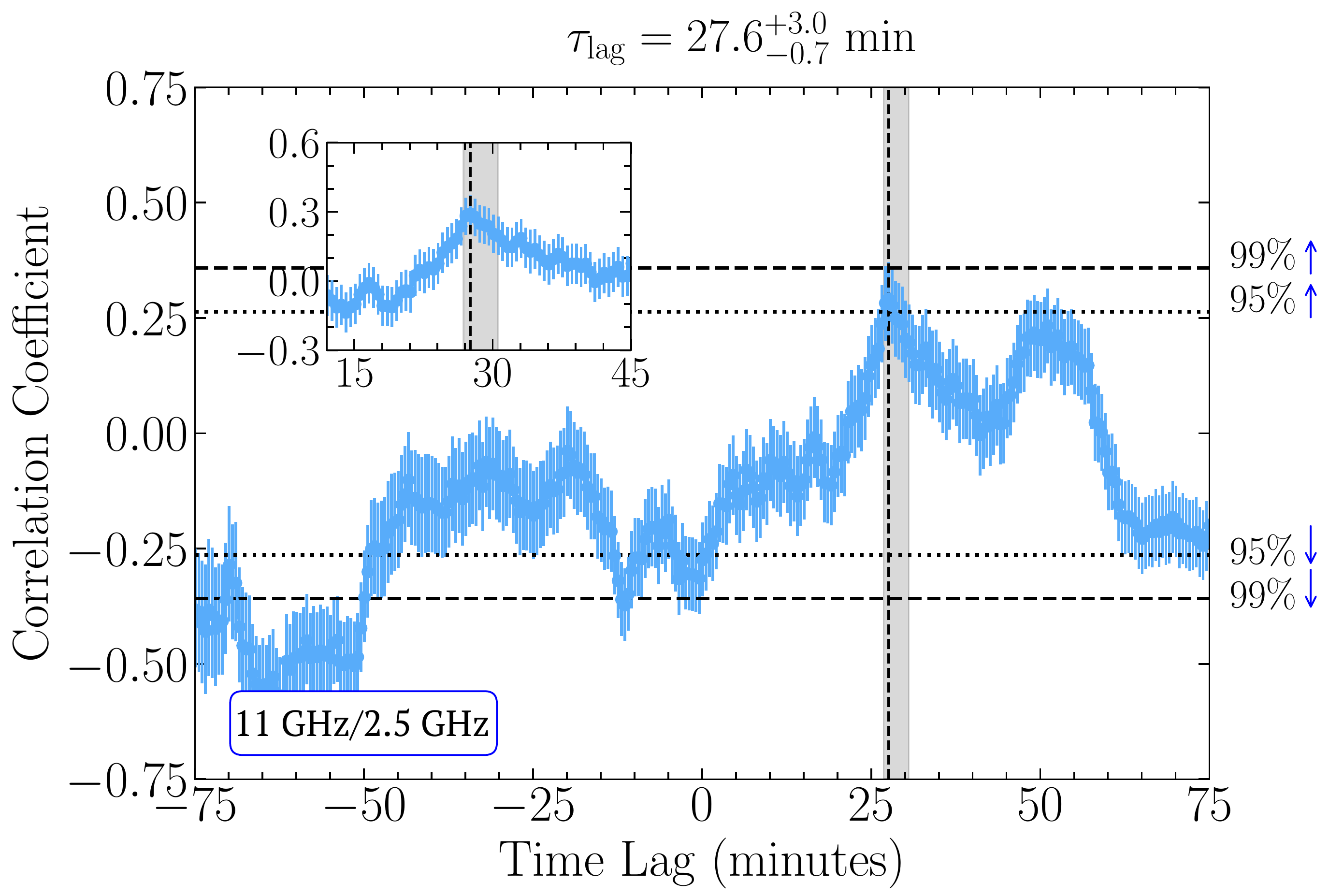}} 
 \caption{\label{fig:ccf} Cross-correlation functions (ZDCFs) between X-ray and radio emission from Cyg X-1; between 3--79 keV/11 GHz ({\sl top left}), 11 GHz/9 GHz ({\sl top right}), 11 GHz/3.5 GHz ({\sl bottom left}), and 11 GHz/2.5 GHz ({\sl bottom right}). The peak of each CCF (indicating the strongest positive correlation at $\tau_{\rm lag}$) is shown by the black dotted line, with the 68\% fiducial confidence interval indicated by the shaded grey region, and 99/95\% significance levels indicated by the black dashed/dotted lines (see \S\ref{sec:ccf} for details). The insets in each panel display a zoomed view of the region surrounding the peak of each CCF. A positive time lag indicates that the lower frequency band lags behind the higher frequency band. Through taking a Bayesian approach (whereby negative lags are considered unphysical; see \S\ref{sec:ccf}), we measure time lags between the X-ray and radio bands, as well as between the individual radio bands, ranging from minutes to tens of minutes.
 }
 \end{figure*}

Simultaneous radio and X-ray frequency light curves of Cyg X-1 taken with the VLA and {\em NuSTAR} are shown in Figure~\ref{fig:cyglc} (also see Figure~\ref{fig:Xlc}). In the radio light curves, the flux density ranges from 12--19 mJy, with the average flux density at all frequencies $\sim 15$ mJy (as expected from historical observations of the source; \citealt{poo99cyg,fen2000,bro02}). However, we also observe structured variability in the radio light curves, in the form of small amplitude flaring events, on top of smoother, longer-timescale variations. {For example, the largest flare detected at 11 GHz ($\sim$ 20:06 UT or 4000 sec in Figure~\ref{fig:cyglc}) reaches an amplitude of $\sim 30 \%$ of the average flux density over a timescale of $\sim$ 12 min (corresponding to a brightness temperature of $\sim 10^9$ K, consistent with other synchrotron flare events from XBs; \citealt{pie15}}). This large flare is asymmetric in shape, showing a secondary peak $\sim 30$ min after the main peak.
Similarly, in the X-ray light curves we also observe structured variability, including a couple of small flares {(e.g., at $\sim$ 19:25 UT or 2000 sec in Figure~\ref{fig:cyglc})}, all of which precede the radio flaring activity.

Through comparing the emission we observe in the different radio bands, the lower radio frequency emission appears to lag the higher radio frequency emission (with the lag increasing as the radio frequency decreases), and the variability in the 2.5/3.5 GHz light curves appears to be of lower amplitude when compared to the 9/11 GHz light curves. Additionally, the lack of clear flares (with well-defined rise and decay phases) in the 2.5/3.5 GHz light curves suggests the variability in these lower frequency radio bands may be more smoothed out than in the higher frequency radio bands. This emission pattern 
is consistent with what we expect from a compact jet, where the higher radio frequency emission originates in a region with a smaller cross-section closer to the black hole, while the lower radio frequencies probe emission from larger regions further downstream in the jet flow.

\subsection{Cross-Correlation Functions}
\label{sec:ccf}
The morphology of our Cyg X-1 light curves hints at a potential correlation between the emission within the radio bands, and between the radio and X-ray bands.
To characterize any correlations and place estimates on time-lags between the bands, we computed cross-correlation functions (CCFs) of our light curves using the z-transformed discrete correlation function (ZDCF; \citealt{alex97,alex13a}). The ZDCF algorithm is designed for analysis of unevenly sampled light curves, improving upon the classic discrete correlation function (DCF; \citealt{ed88}) or the interpolation method \citep{gask87}. The calculated CCFs are displayed in Figure~\ref{fig:ccf}, {where we use the full radio light curves to create the CCFs, not just the continuous data chunks (as is done for the Fourier analysis in \S\ref{sec:fa}), as the ZDCF algorithm can handle unevenly sampled data.}

The location of the CCF peak will indicate the strongest positive correlation, and thus the best estimate of any time-lag between the light curves from different observational frequency bands. To estimate the CCF peak with corresponding uncertainties, we implement the maximum likelihood method of \citet{alex13a}. We note that this method estimates a fiducial interval rather than the traditional confidence interval. The approach taken here is similar to Bayesian statistics, where the normalized likelihood function (i.e., fiducial distribution) is interpreted as expressing the degree of belief in the estimated parameter, and the 68\% interval around the likelihood function's maximum represents the fiducial interval. 
Additionally, to estimate the significance level of any peak in the CCF, we perform a set of simulations
allowing us to quantify the probability of false detections in our CCFs, by accounting for stochastic fluctuations and intrinsic, uncorrelated variability within each light curve.
For these simulations, we randomize each light curve 1000 times (i.e., Fourier transform the light curves, randomize the phases, then inverse Fourier transform back, to create simulated light curves that share the same power spectra as the real light curves) and calculate the CCF for each randomized case. We then determine the 95\% and 99\%  significance levels based on the fraction of simulated CCF data points (at any lag) above a certain level.

Our calculated CCFs (Figure~\ref{fig:ccf}) suggest the presence of a correlation between the X-ray and radio emission from Cyg X-1, as well as between the emission in the individual radio bands. In particular, we measure a time lag between the 3--79 keV X-ray and the 11 GHz radio bands of $33.5^{+1.9}_{-1.7}$ min, and time lags between the 11 GHz and 9, 3.5, 2.5 GHz radio bands of $1.0^{+0.9}_{-0.3}$ min, $24.1^{+1.6}_{-1.3}$ min, and $27.6^{+3.0}_{-0.7}$ min, respectively (although see below for caveats on the 11/3.5 GHz and 11/2.5 GHz lag measurements). 
With these lags, we observe a trend with observing frequency, where the lower frequency bands always lag the higher frequency bands (and the lag increases as the observing frequency decreases in the comparison band). This trend is consistent with what we expect from emission originating in a compact jet, where the lags presumably trace the propagation of material downstream along the jet flow (from higher frequencies to lower frequencies; \citealt{mal03,ganh08,cas10,gan17}). We also split the 8--12 GHz radio data into four sub-bands (centered on 8.75, 9.25, 10.75 and 11.25 GHz) and re-ran our CCF analysis. However, the measured lags we obtain are all consistent with each other within uncertainties. Thus we do not gain any new information through splitting our radio data into finer frequency bands, and do not report on these data hereafter.

We note that when comparing the 3--79 keV/11 GHz and 11 GHz/9 GHz bands, the CCFs show relatively symmetric peaks at the measured time lag, both of which reach or exceed the 99\% significance level. Therefore, we consider these detected time lags statistically significant, and are confident they are tracking a real correlation between the light curves.  On the other hand, when comparing the 11 GHz/3.5 GHz and 11 GHz/2.5 GHz bands, the CCFs display much more complicated structure, including secondary peaks and anti-correlation dips at negative lags.
While the measured time lag peaks represent statistically significant correlations (i.e., reaching or exceeding the 99\% significance level), the secondary peaks can at times reach a lower, but still significant level (e.g., peak at $\sim10$ min when comparing 11 GHz/3.5 GHz, or peak at $\sim50$ min when comparing 11 GHz/2.5 GHz), and the anti-correlation dips can reach levels more significant than our positive lag peaks. 
{Therefore, even though the measured 11 GHz/3.5 GHz and 11 GHz/2.5 GHz radio lags are nominally statistically significant, and are physically plausible in terms of what we may expect from compact jet emission, we are unable to assign them as much credence  as the lags involving the higher frequency radio bands. Nonetheless}, for the remainder of this paper we opt to take a Bayesian approach, whereby we consider the negative lags as unphysical, and take our measured lags to be the best estimate of the true lags between these radio bands. The reader should keep these caveats in mind when considering the 11/2.5 GHz and 11/3.5 GHz lag interpretations moving forward.
Lastly, an important technical caveat to note in our CCF calculations is that we are calculating the CCFs up to delays that are comparable to the duration of the data set (i.e., outside the stationarity limit). As the CCF is formally defined only on the assumption of stationarity, we are pushing the CCF method to its limits. We thus choose to remain conservative in our claims of time lag detections (especially for the lags calculated from the lower frequency 2.5/3.5 GHz radio bands, which are not clearly seen in the light curves themselves) in this paper given these statistical limitations of our methods.

In the literature, there exist a few studies reporting on X-ray/radio correlations on short (minute) timescales in Cyg X-1. In particular, \citet{gleiss4} found no statistical evidence for correlations between X-ray and radio emission in Cyg X-1 on timescales $<5$ hours, suggesting that any possible correlations they observed were consistent with artifacts of white noise statistics. The significance level simulations we performed in this paper rule out our correlations being statistical artifacts. Further, the radio frequency data used in the \citet{gleiss4} study was obtained with the Ryle Telescope\footnote{The Ryle Telescope has now been decommissioned, but some of its antennas are in continued use with the current Arc-Minute MicroKelvin Imager Large Array (AMI-LA).}, which was much less sensitive than the current VLA (which has more antennas with larger dishes and wider bandwidth). Therefore, our study is able to measure lower-level signals, giving us better statistics in our CCFs when compared to the \citet{gleiss4} study. Additionally, \citet{wil07} report the detection of an X-ray/15 GHz lag of $\sim7$ min, which is significantly shorter than the measured $\sim 33$ min X-ray/11 GHz lag reported in this work. However, we note that the \cite{wil07} data sampled Cyg X-1 during a transition from the soft to hard accretion state, and displayed different flare morphology (e.g., a higher amplitude of $\sim4$ times the average flux level and a much more symmetric radio flare shape) when compared to our data. The jet emission in BHXBs during accretion state transitions is often dominated by emission from discrete jet ejections (typically characterized by higher amplitude, more symmetric radio flares; e.g., \citealt{tetarenkoa17}), rather than a compact jet (as seen in the hard state). Therefore, the difference between these measured time lags could be due to both studies sampling a different form of jet emission with different properties (e.g., jet size scales, bulk speeds, opening angles, energetics). 
Although, we note that it is the hard to soft accretion state transitions (rather than the soft to hard transitions) that often show jet ejecta, and these jet ejecta have only been detected once in Cyg X-1 \citep{fenstir06}.
Alternatively, in either study it is possible that the pairs of correlated X-ray/radio flares were misidentified, and are in turn not actually correlated (e.g., it is possible that additional X-ray flares occur within the orbital gaps in the data; see for example \citet{cap17} who show it can be difficult at times to know whether X-ray/radio flares are truly connected).

\subsection{Fourier Analyses}
\label{sec:fa}
To better characterize the variability properties of our radio frequency light curves of Cyg X-1, we opted to also perform Fourier domain analyses. As part of these analyses, we calculate the power spectral density (PSD), as well as perform cross-spectral analyses, allowing us to characterize lag and correlation behaviour across many distinct timescales of variability \citep{vau97,utt15}. While Fourier domain analysis is common practice at higher frequencies (infrared, optical, X-ray; e.g.\  \citealt{ganh08,cas10,vinc18}), previous studies of this nature in the radio bands are limited (e.g., see \citealt{nip05}, probing timescales as short as days only). Here we present the first Fourier domain study of a BHXB (including a cross-spectral analysis) undertaken at radio frequencies on timescales as short as seconds. We use the {\sc stingray} software package\footnote{\url{https://stingray.readthedocs.io/en/latest/}} \citep{stingray} for all of our Fourier domain analyses below.

\begin{figure}
   \centering
   
   \subfloat{ \includegraphics[width=0.9\columnwidth,height=5cm]{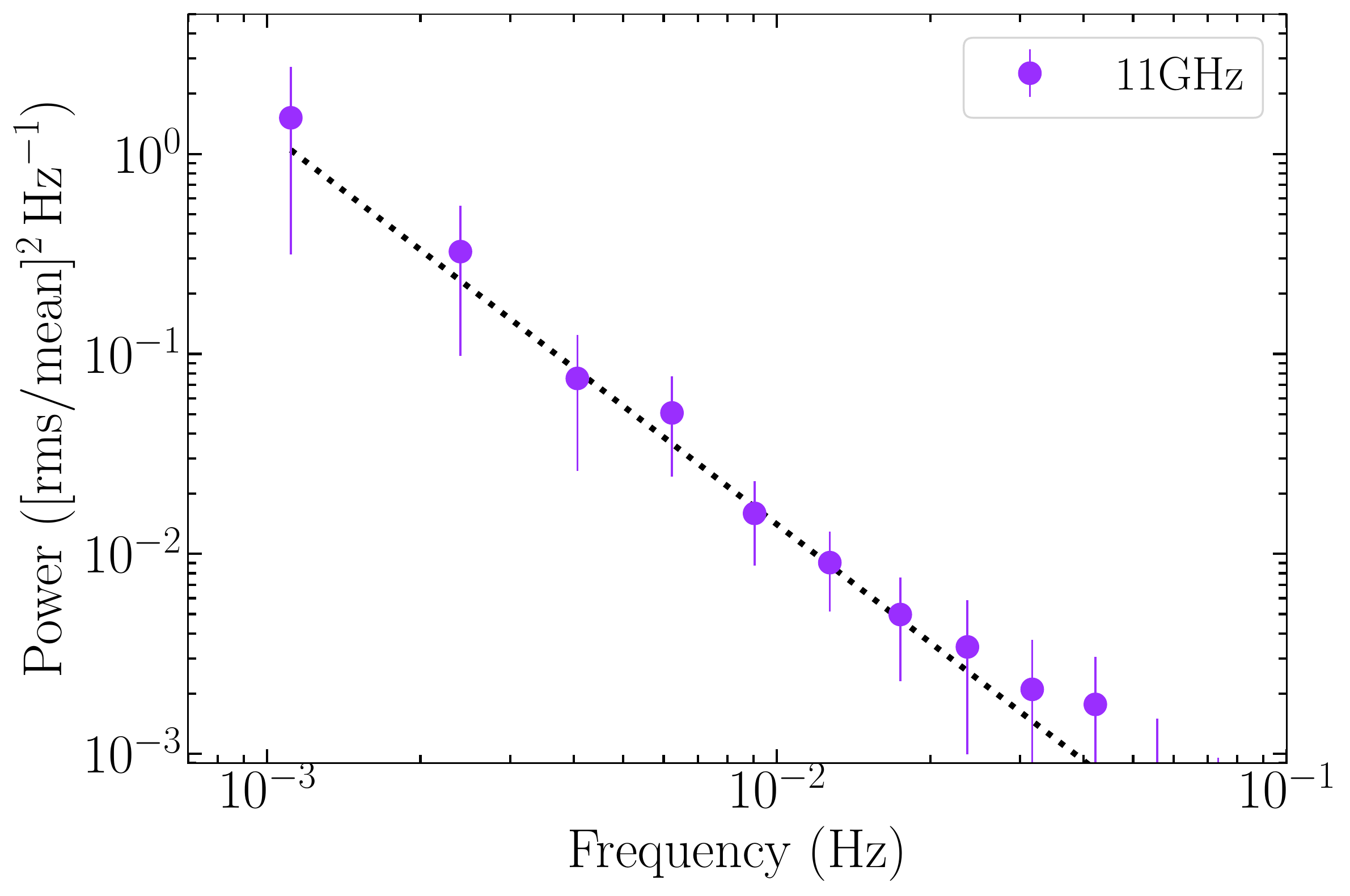}}\\[-0.45cm]
   \subfloat{ \includegraphics[width=0.9\columnwidth,height=5cm]{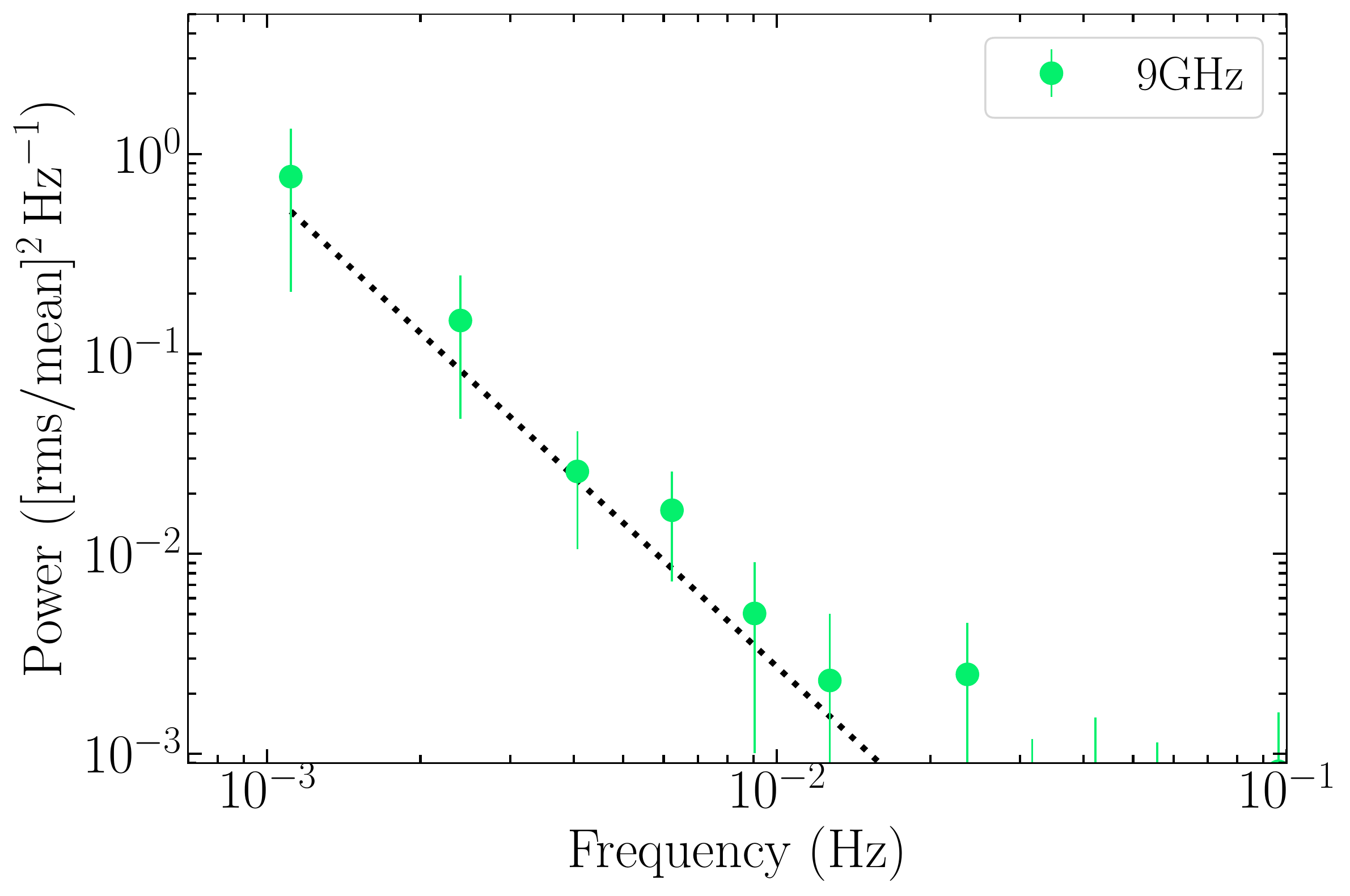}}\\[-0.45cm]
   \subfloat{ \includegraphics[width=0.9\columnwidth,height=5cm]{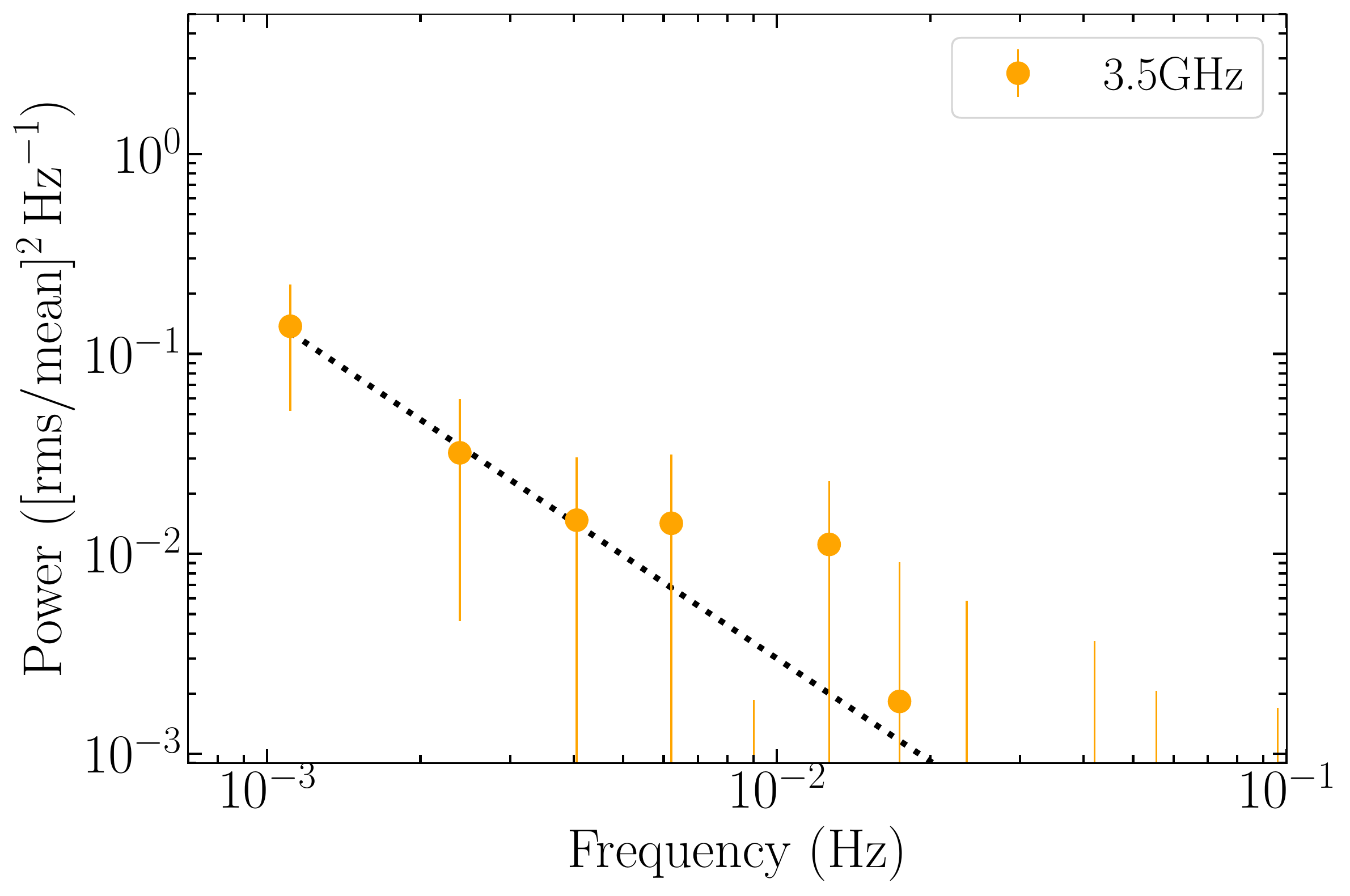}}\\[-0.45cm]
   \subfloat{ \includegraphics[width=0.9\columnwidth,height=5cm]{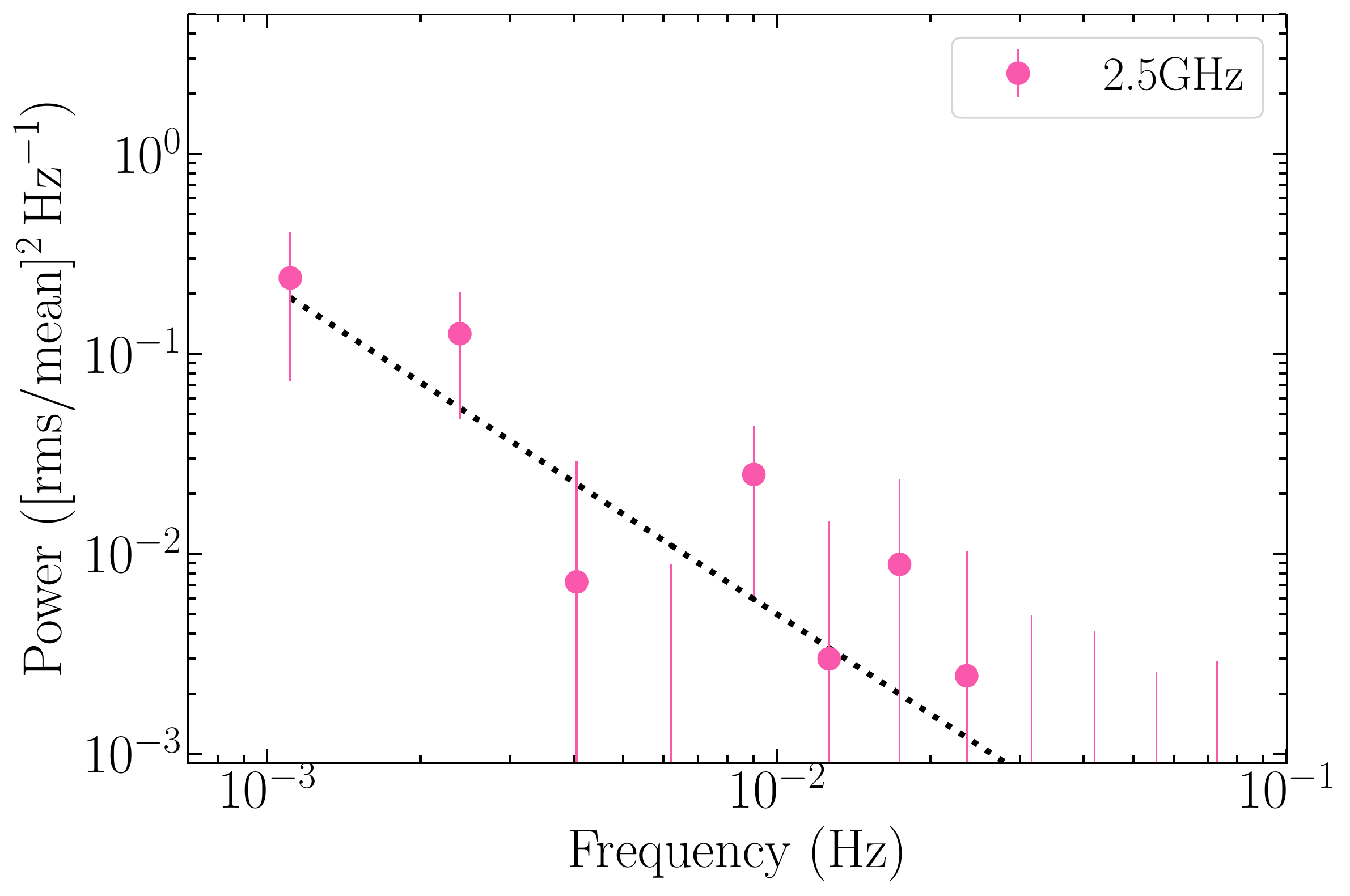}} 
 \caption{\label{fig:psdr} White noise subtracted radio PSDs of Cyg X-1. Top to bottom, the panels display the power spectra from the 11, 9, 3.5, and 2.5 GHz bands. The PSDs in the 9/11 GHz bands are created from data imaged with 1 sec time-bins, while the 2.5/3.5 GHz PSDs are created from data imaged with 5 sec time-bins. {The black dotted line represents the single power-law fits to the PSDs (see \S\ref{sec:psd} for details)}. While all the radio bands display a similar PSD shape (decreasing power at higher Fourier frequencies), when considering the longest timescales, we observe higher power in the higher frequency radio band PSDs. Note that the pre-white noise subtracted PSDs are shown in Figure~\ref{fig:psdr_wn}.
 }
 \end{figure}

\subsubsection{Power Spectra}
\label{sec:psd}
\begin{figure}
   \centering
   {\includegraphics[width=1\columnwidth]{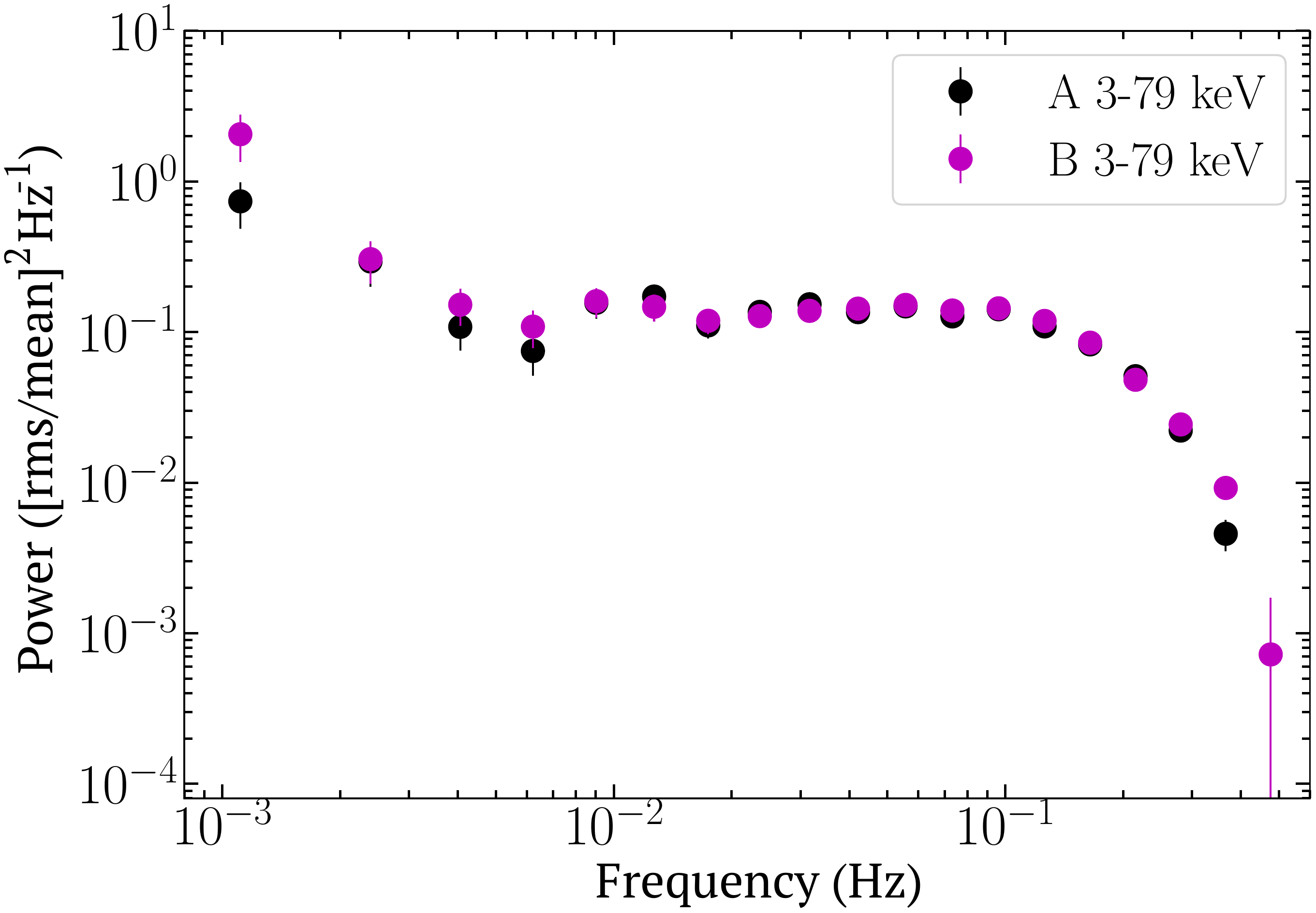}}
 \caption{\label{fig:psdx} White noise subtracted X-ray PSDs of Cyg X-1 in the 3--79 keV band. These power spectra are created from data with 1 sec time-bins (matching our radio frequency light curves). PSDs for the two {\em NuSTAR} modules are shown separately (FPMA represented by the black markers, FPMB represented by the magenta markers). The X-ray PSDs show a clear turnover at higher frequencies ($\sim0.3$ Hz), and no quasi-periodic oscillations are observed.
 }
 \end{figure}

To create our radio PSDs, we consider only the continuous chunks of data in each radio light curve (i.e., first $\sim$75 min at 9/11 GHz and second $\sim$75 min at 2.5/3.5 GHz), and divide the light curves into 15 min segments,
averaging them to obtain the final PSD. The segment size was chosen to reduce the noise in the PSDs. 
Further, a geometric re-binning in frequency was applied (factor of $f=0.3$, where each bin-size is $1+f$ times larger than the previous bin size) to reduce the scatter at higher Fourier frequencies. All our PSDs are normalized using the fractional rms-squared formalism \citep{bel90} and white noise has been subtracted (white noise levels were estimated by fitting a constant to Fourier frequencies above 0.05 Hz \& 0.01 Hz for the 9/11 GHz \& 2.5/3.5 GHz bands, respectively; see Appendix \ref{sec:app2}).
Note that PSDs in the 9/11 GHz bands are created from data imaged with 1 sec time-bins, while the 2.5/3.5 GHz PSDs are created from data imaged with 5 sec time-bins\footnote{The maximum frequency for Fourier analysis is the Nyquist frequency of $\nu_{\rm N}=\frac{1}{2}\,{t_{\rm res}}^{-1}=0.5\, {\rm and}\, 0.1$ Hz for the 9/11 GHz and 2.5/3.5 GHz PSDs, respectively ({$t_{\rm res}$ represents the time resolution of our light curves.)}}. We use a different imaging timescale between radio bands to roughly match the rms noise in each time-bin's image across the bands, while still leaving enough of a Fourier frequency lever arm to accurately measure the white noise level.
To create our X-ray PSDs, we follow the same averaging and binning procedure as the radio PSDs (additionally making use of the good time interval feature\footnote{This good time interval (GTI) feature computes the start/stop times of equal time intervals, taking into account both the segment size (timescale over which the individual PSDs are averaged together) and the GTIs. This ``time mask" (array of time stamps) is then used to start each FFT from the start of a GTI, and stop before the next gap in the data (end of GTI).} of the {\sc stingray} package to take into account the orbital gaps in the data) and use frequencies above 0.4 Hz to estimate the white noise level.
Figure~\ref{fig:psdr} displays the radio PSDs, while Figure~\ref{fig:psdx} displays the X-ray PSDs.

\begin{figure*}
   \centering
   
   \subfloat{ \includegraphics[width=1.95\columnwidth]{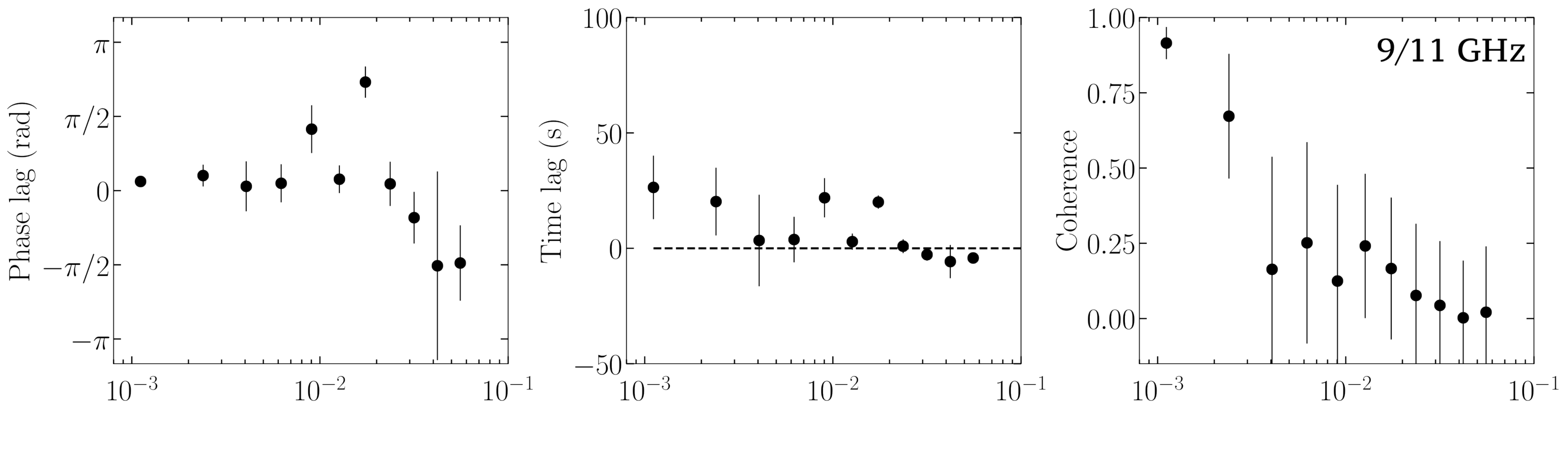}}\\[-0.6cm]
   \subfloat{ \includegraphics[width=1.95\columnwidth]{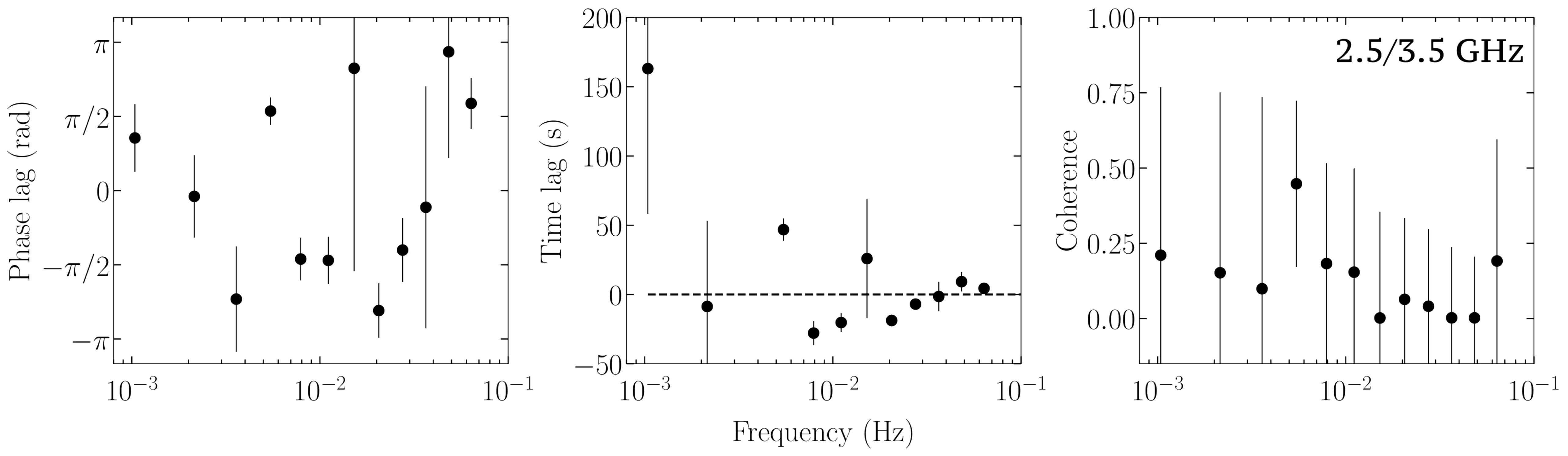}}
 \caption{\label{fig:csdr} Cross-spectral analysis of radio emission from Cyg X-1. The {\sl top} three panels display a cross-spectral analysis between the 9/11 GHz bands (using 1 sec time-bin data), while the {\sl bottom} three panels display a cross-spectral analysis between 2.5/3.5 GHz bands (using 10 sec time-bin data, to further increase the signal to noise when compared to the 5 sec time-bin data used in the PSD procedure for these bands). Left to right, the panels display phase lags, time lags, and coherence. Positive lags indicate that the lower radio frequency band lags the higher radio frequency band, and a time lag of zero is indicated by the black dashed line in the {\sl middle} panels. The higher frequency radio bands ({\sl top}) show high levels of coherence and near constant time-lags on the longer timescales (Fourier frequencies $<0.01$ Hz), while the lower frequency radio bands ({\sl bottom)} comparatively show very little coherence across all time scales.
 }
 \end{figure*}

The radio PSDs all appear to display a power-law type shape, where the highest power occurs at the lowest Fourier frequencies (corresponding to the longest timescales sampled) and no significant power is observed above Fourier frequencies of $\sim 0.03$ Hz ($\sim 30$ sec timescales). {Through fitting a single power-law to the radio PSDs, we find power-law indices of $-1.96^{+0.29}_{-0.31}$, $-2.39^{+0.49}_{-0.50}$, $-1.71^{+0.51}_{-0.74}$, $-1.66^{+0.85}_{-0.66}$ for the 11, 9, 3.5, and 2.5 GHz radio bands, respectively. We tested whether these slopes could be due to leakage (i.e., where variability on timescales longer than the interval over which our PSD is calculated adds "fake" power into the PSD) by de-trending our light curves prior to redoing the PSD calculation. Through these tests we confirm that the slopes we measure are not a result of leakage. The power-law indices we observe in our PSD (indices of $\sim 2$) could be produced as a result of red noise, or we may be observing the high Fourier frequency part of a band-limited noise feature (which is commonly observed at other wave-lengths in XBs).} 
Additionally, we observe a trend where the PSDs from the higher frequency radio bands display larger power on long timescales than the lower frequency radio bands (where the larger uncertainties make it difficult to compare the power between the radio bands on smaller timescales/higher Fourier frequencies). 

These observed PSD features fit with our compact jet picture, where we expect higher radio frequencies to display larger variability amplitudes and the highest Fourier frequency variations to be more suppressed, due to the increasing size scale of the emitting region at radio frequencies (when compared to the X-ray) smearing the variability signal. 
{In fact, size scale constraints from VLBI imaging of Cyg X-1 \citep{stir01} suggest jet cross-sections which are consistent with the light crossing times of the shortest timescales for which we detect significant power (in $\sim 30$ sec a signal propagating at the speed of light travels $1\times10^{12}$ cm). In particular, for a conical jet, the jet cross-section at distance $z_0$ from the jet launching region is represented by $z_{\rm cross}=z_0 \tan{\phi}$ (where $\phi$ is the jet opening angle). VLBI imaging studies of Cyg X-1 observe the jet out to $\sim15$ mas scales, along a jet axis inclined to our line of sight by $\sim27$ degrees \citep{or11}, corresponding to $z_0\sim8.9\times10^{14}$ cm at 8.4 GHz. With an opening angle of $<2$ degrees, we in turn estimate a cross-section at 8.4 GHz of $<3.1\times10^{13}$ cm}. {Further, given a brightness temperature limit of $10^{12}$ K for synchrotron emission, the shortest timescale over which we would expect to be able to measure significant variability (i.e., at least 0.5--1\% of the average flux density, given the variability amplitudes seen in our radio PSDs) with the VLA would be on the order of seconds.}

The X-ray PSDs also appear to display a power-law type shape at lower Fourier frequencies ($<0.01$ Hz), then flatten out at higher Fourier frequencies, with a turnover occurring at $\sim0.3$ Hz. No quasi-periodic oscillations are observed, although we do observe a bump around $\sim 0.008$ Hz (this X-ray PSD shape is consistent with previous studies; e.g., \citealt{now99b}). Interestingly, when comparing the X-ray and radio PSDs, we find that the X-ray PSDs show similar power at the lowest Fourier frequencies (longest timescales) as the highest frequency radio band PSDs. This feature was also seen at Fourier frequencies as low as $\sim10^{-6}-10^{-7}$ Hz by \citet{nip05} in their earlier, longer timescale study (and notably was a unique feature to Cyg X-1, not observed in the long timescale radio PSDs of other XB sources; GRS 1915+105, Cyg X-3, and Sco X-1).
However, the X-ray emission is significantly more variable overall, displaying an integrated fractional rms (across the 0.001 to 0.5 Hz range) of $\sim 20$\%, while all the radio bands show $\sim 2-7$\%.

\subsubsection{Cross Spectra}
Through cross-spectral analyses we can examine the causal link between two time series signals.
For this analysis we only consider the segments of the respective radio light curves for which we have continuous data with no scan gaps (as was done in creating our PSDs) and use the same averaging/binning procedure as described for the PSDs.
Further, we only compare the signals we observe between the two-basebands in each VLA frequency band (i.e., between 9 and 11 GHz, and between 2.5 and 3.5 GHz), as our observational setup only allowed for the simultaneous, continuous observations needed for such an analysis between these radio frequencies.

Figure~\ref{fig:csdr} displays the results of our cross spectral analyses, where we show three different metrics used to quantify the causal relationship between the two time-series signals: phase lags, time lags, and coherence. The lags describe the phase/time differences between intensity fluctuations for each Fourier frequency component, while the coherence is a measure of the fraction of the rms amplitude of one signal (at a given Fourier frequency) that can be predicted from the second signal through a linear transform (i.e., the degree of linear correlation between the two signals as a function of Fourier frequency).

When considering the higher frequency radio bands (9/11 GHz), we observe a high level of coherence on longer timescales ($f<0.005$ Hz), which drops to $\sim 0.25$ at $\sim0.005$ Hz, then to a point where no significant correlation is detected above $\sim0.05$ Hz. Additionally, we observe a relatively constant trend in the phase lags, corresponding to a mostly constant time lag of a few tens of seconds, across Fourier frequencies up to $\sim$0.01 Hz (although we do note that there is some scatter in the time-lags at $\sim0.005$ Hz). These time lags are consistent with the lower end of the confidence interval estimated from our CCF measured lag of $1.0^{+0.9}_{-0.3}$ min, within the uncertainty limits. However, when considering the lower frequency radio bands (2.5/3.5 GHz), we observe very little coherence, even on the longer timescales ($\sim 0.25$ at 0.001 Hz, but coherence consistent with zero within the large uncertainties). Further, there is no clear trend in the phase lags, and the time lags are mostly consistent with zero. 
However, despite the large uncertainty, the time lag in the lowest Fourier frequency bin is consistent with what we might expect from our CCF measured lags of $\sim3-4$ min.


\section{Discussion}
\label{sec:discuss}
To characterize the variability properties of the compact jet emission in the BHXB Cyg X-1, we have performed a time domain analysis on multi-band VLA radio and {\em NuSTAR} X-ray observations of this system. We implemented several different metrics in our analysis; cross-correlation functions, PSDs, and cross-spectral analysis. In the following sections, we discuss what each of these metrics reveals about the jet variability properties and how these variability signals propagate down the jet. Additionally, we derive constraints on jet speed, geometry, and size scales.

\subsection{Interpreting PSDs, lags, and coherence}

The X-ray and radio PSDs presented in Figures~\ref{fig:psdr} and \ref{fig:psdx} show remarkably similar power on the longest timescales, but display very different shapes. In particular, the radio bands display a monotonically decreasing trend with frequency (where there is no significant power above Fourier frequencies of $\sim$0.03 Hz), while the X-ray band displays a double power-law type shape that turns over at $\sim$0.3 Hz. In previous BHXB jet variability studies performed at higher frequencies \citep{ganh08,cas10,vinc18}, infrared and optical PSDs of compact jet emission displayed a similar shape to our X-ray PSDs, where a break is observed at higher Fourier frequencies ($\sim3/1$ Hz at optical/infrared bands in GX 339-4). This damping of the power at higher Fourier frequencies is thought to reflect the physical size scale of the emitting region. Therefore, we might expect any break in the radio PSDs to occur at lower Fourier frequencies than we can probe with our data set (e.g., a VLBI transverse size scale estimate of $\sim3.1\times10^{13}$ cm at 8.4 GHz gives a light crossing time of $\sim 1100$ sec or $\sim9\times10^{-4}$ Hz; see Figure 6 in \citealt{mal14} where theoretical jet models also predict this PSD shape). In the Fourier analysis of \citet{nip05}, who probed much longer timescales than our study (days-months), the radio PSD shows a higher power (at all sampled Fourier frequencies) when compared to our PSDs, that is nearly constant at Fourier frequencies between $10^{-6}-10^{-5}$ Hz. Therefore, this is suggestive of a break in the radio PSDs occurring in the un-sampled Fourier frequency range ($10^{-5}-10^{-3}$ Hz). If this is the case, the radio PSDs we present here could be sampling the drop off in power after a lower Fourier frequency turnover. A set of longer radio observations (sampling the $10^{-5}-10^{-3}$ Hz Fourier frequency range) and/or observations at higher radio/sub-mm frequencies are needed to confirm this theory, and test whether there is a frequency dependent trend in the location of this turnover across different radio bands (as suggested between the infrared/optical PSD turnovers observed in GX 339-4). Measuring such a trend could (in principle) be used to map out the jet size scale in different regions of the jet flow.

In our cross-spectral analysis, we observed that the higher frequency radio bands (9/11 GHz) were highly correlated on longer timescales ($>200$ sec, $f<0.005$ Hz), and display a roughly constant lag with Fourier frequency consistent with our CCF measured lag. However, in the lower frequency radio bands (2.5/3.5 GHz) we observed very little correlation across all timescales, and a lag consistent with our CCF analysis only in the lowest Fourier frequency bin. In the situation where one signal is related to the other through a delay, due to propagation of variability down the jet flow, we would expect both the coherence and time lags to be constant with Fourier frequency. While we observe constant time lags below a certain Fourier frequency in our analysis, the coherence either drops at higher Fourier frequencies (9/11 GHz) or is very low across all Fourier frequencies (2.5/3.5 GHz). Therefore, if the radio signals are related by a propagation type model, some other process must be damping the correlation at higher Fourier frequencies. In our radio signals, the coherence and radio variability amplitude (traced through the PSDs) appear to be damped over the same range of Fourier frequencies. {Therefore, it seems plausible that the loss of coherence we observe may be due to processes occurring on timescales shorter than the propagation time between the regions emitting our radio signals, such as turbulence in the jet flow \citep{gleiss4}, distorting the variability signals.}
In BHXB jet variability studies performed at higher infrared frequencies \citep{vinc18}, a similar trend (constant time lags with Fourier frequency but a decreasing trend in coherence) was seen in the cross-spectral analysis. However, in the infrared bands there appears to be a different mechanism at work (when compared to our radio signals) that reduces the strength of the correlation between signals, but does not affect the infrared variability amplitude (i.e., the PSD breaks at 1 Hz but the coherence begins to drop at lower Fourier frequencies).

Other factors could also be contributing to the loss of coherence between the radio signals.
In particular, a strong noise component in the radio light curves may lead to a loss of coherence (as random noise signals are not correlated). 
To further investigate this possibility, we opted to compute the intrinsic coherence (using Equation 8 in \citealt{vau97}), which takes into account noisy signals by applying a correction term to the measured coherence. Through this computation, we find that the intrinsic coherence is nearly identical to the measured coherence at the Fourier frequencies for which the \citet{vau97} expression is valid ($<0.01$ Hz). This suggests that the noise component in the radio light curves is not a significant cause of any loss of coherence we observe. However, we note that both the measured/intrinsic coherence measurements between the 2.5/3.5 GHz radio signals display large uncertainties. Therefore, the noise component limits our ability to accurately calculate and compare the intrinsic/measured coherence between the lower frequency (2.5/3.5 GHz) radio signals.
Additionally, phase wrapping in the cross-spectra could also be a contributing factor in causing the loss of coherence and lack of flat time-lags at higher Fourier frequencies. The effect of phase wrapping is likely not an issue for the 9/11 GHz cross-spectra, as the CCF predicted delay of $\sim1$ min/0.01 Hz between these bands occurs at higher Fourier frequencies than the point where the coherence and time-lags drop out. However, this phase wrapping effect may be relevant for the 2.5/3.5 GHz cross-spectra, where the CCF predicted delays ($\sim3-4$ min/$4-5\times10^{-3}$ Hz) between these bands match closely with the Fourier frequencies where the coherence and time-lags drop out. However, as the coherence is already small at these lower Fourier frequencies in the 2.5/3.5 GHz cross-spectra, phase wrapping can not be the only explanation for the coherence and time-lag drop out.

Further, a loss of coherence could also occur if more than one source of emission contributes to the signals in these bands (even if individual sources produce coherent variability; \citealt{vau97}) or if the signals are acted on by a nonlinear process (e.g., internal shocks in the jet flow). For instance, the strong stellar wind from the companion star in Cyg X-1 has been shown to partially absorb the radio emission by up to about 10 \% (\citealt{poo99cyg,bro02}), and thus could potentially be distorting the radio signals we observe. While \citet{bro02} showed that 2.25 GHz radio emission does not vary much around the orbit, implying that this lower frequency emission could originate outside the wind photosphere (see Figure 1 in \citealt{bro02}), our observations were taken at an orbital phase where the black hole was behind a significant amount of the wind (orbital phase 0.88, where superior conjunction is at orbital phase 0). Therefore, we would expect the wind to have close to a maximal effect on the radio emission we observe during our observations. Further, \citet{bro02} estimate the size of the wind photosphere to be
$4.6\times10^{14}$/$1.8\times10^{14}$ cm at 2.25/8.3 GHz, which are both larger than the size scales to the radio emitting regions estimated from our CCF X-ray to radio lag estimates, suggesting that the stellar wind could reasonably be affecting our radio signals. {We note that the radio emission originating from the stellar wind itself is likely to display much smaller radio flux densities when compared to the compact jet. For example, using the work of \citealt{wrbar75} \& \citealt{leith95} (Equation 1 of \citealt{leith95} with stellar wind parameters given in \citealt{bro02}), we estimate the stellar wind from the Cyg X-1 companion star produces a flux density of $\sim 0.03/0.08$ mJy at 2.5 GHz/11 GHz.}

Additionally, the emission from the counter-jet (i.e., the portion of the bi-polar jet travelling away from the observer) could also be contaminating our radio signals. The ratio of the flux densities between the approaching and receding jets in Cyg X-1 can be estimated 
using $\frac{F_{\rm app}}{F_{\rm rec}}=\left( \frac{1+\beta\,\cos{\theta}}{1-\beta\,\cos{\theta}} \right)^{2-\alpha}$ (where $\beta$, $\theta$, and $\alpha$ represent the jet speed, inclination angle of the jet axis to the line of sight, and the radio spectral index, respectively). However, given values of $\beta=0.92$, $\theta=27$ degrees, and $\alpha=0$ (see \S\ref{sec:speed}), we find $\frac{F_{\rm app}}{F_{\rm rec}}\sim102$, suggesting that any contributions from the counter-jet will likely be negligible in our radio signals {(although if the inclination of the jet axis to our line of sight is closer to 40 degrees, as suggested by X-ray reflection modelling, rather than 27 degrees found by \citealt{or11}, this ratio becomes much smaller at $\sim 30$.)}.
Further, the emission in the radio frequency bands could have contributions from a region where the jet is colliding  with the surrounding ISM. These regions can display a working surface shock at the impact point (e.g., \citealt{co05,millerjones11,yang11,rush17}) and typically show brighter flux densities at lower radio frequencies, potentially contributing more to the overall observed emission in lower radio frequency bands (where we see increased loss of coherence). In the case of Cyg X-1, the jet is thought to have carved out a parsec scale cavity in the intervening medium ($\sim2\times10^{5}$ AU; \citealt{gallo05,rus07,sell15}), which may make further jet impact sites closer (less than the VLA beam of 7 arcsec or $10^{4}$ AU at 1.86 kpc) to the central source unlikely. However, we note that Cyg X-1 was in a soft accretion state (with presumably no jet) from 2010 to a few months prior to our observations in early 2016. Therefore, the strong stellar wind from the companion star could have had time during this soft state to refill at least a portion of this cavity, and in fact the jets we observe could be interacting with this newly deposited material (e.g., see \citealt{kolj18}).


\subsection{Constraints on jet properties}
\label{sec:speed}

If the X-ray emission we observe originates in a region close to the black hole (inner accretion flow or at the base of the jet) and information from X-ray emission regions propagates down the axis of the jet to the radio emission regions, then the distance between the radio emission regions and the black hole, $z_{{\rm radio}}$, can be represented as, 
\begin{equation}
z_{{\rm radio}}=\beta\,c\,\tau_{\rm lag} \,(1-\beta\cos\theta)^{-1}\,.
\label{eq:z0}
\end{equation}
Here, $\beta$ represents the bulk jet speed (in units of $v/c$, where $c$ indicates the speed of light), $\tau_{\rm lag}$ represents the X-ray to radio time lag, and $\theta$ represents the inclination angle of the jet axis to our line of sight. The term in parentheses indicates a correction factor due to the transverse Doppler effect (where the interval between the reception of two photons by the observer is smaller than the interval between their emission). {We note that in the case where the X-ray emission originates from an accretion flow that is not co-spatial with the jet base, we may expect some additional time delay in getting the accreted material into the jet base, and subsequently accelerating this material. While these timescales are not known for BHXBs, if for example the jet becomes radiative (i.e., acceleration zone) at $\sim1000$ gravitational radii from the black hole \citep{gan17}, this timescale could correspond to a light travel time delay of tens of milli-seconds. However, in this case, milli-second timescales are minuscule when compared to the tens of minutes delays we are discussing here. Further, if the X-ray variations originate in the accretion disc, we may expect an additional delay for these variations to propagate inwards.}

As we measured time lags between the X-ray and multiple radio bands, we can use Equation~\ref{eq:z0} to solve for the jet speed {by substituting in a metric to relate jet size scale to the radio frequency band. While simple jet models predict $z\propto1/\nu$ \citep{blandford79}, we allow for a more general case in our analysis, where $z\propto1/\nu^\epsilon$.} Through rearranging Equation~\ref{eq:z0}, we obtain,

\begin{figure}
   \centering
   
   { \includegraphics[width=1\columnwidth]{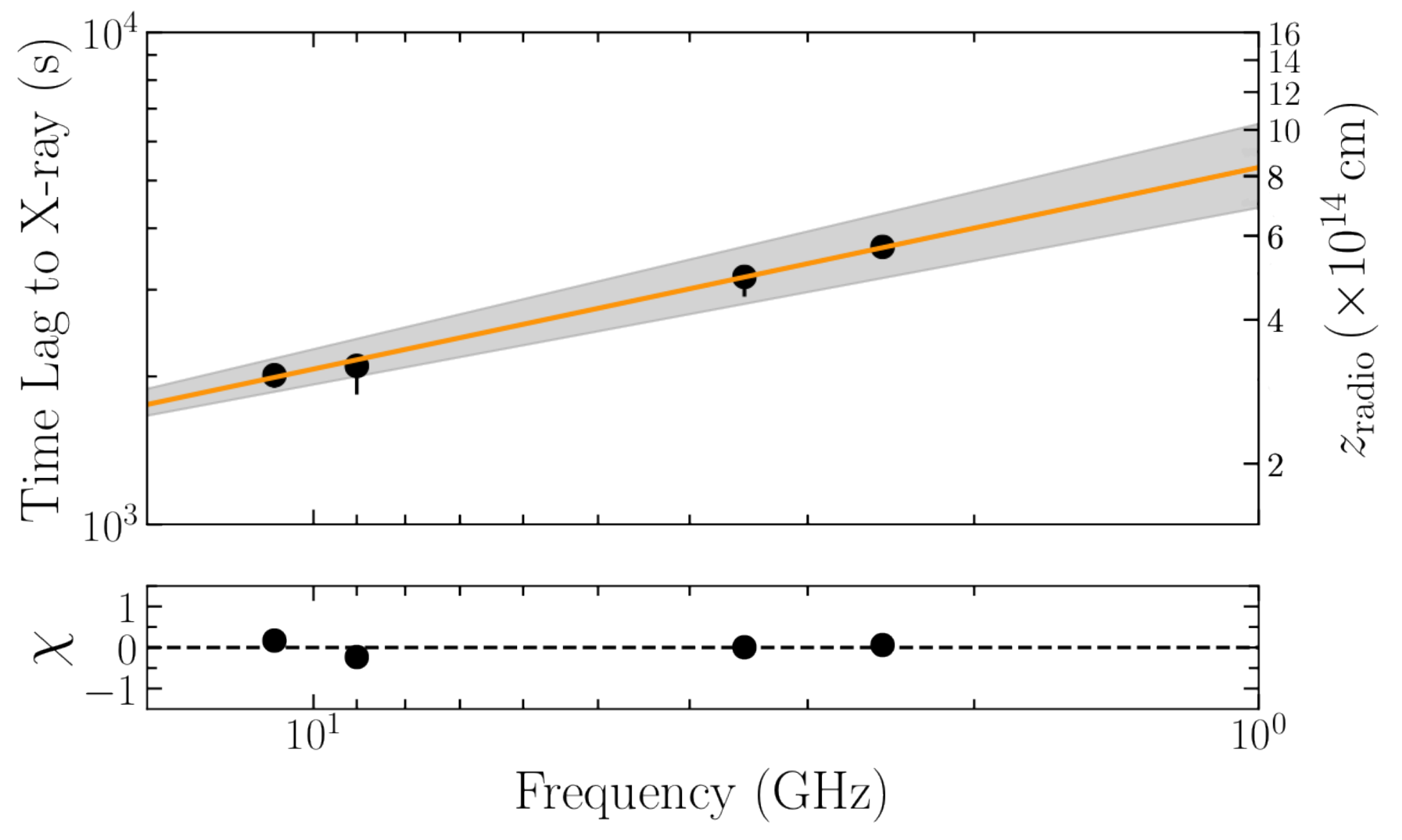}}\\
 \caption{\label{fig:lag_fit} Time lags between the radio and X-ray bands, as a function of radio frequency. The {\sl top} panel displays the time lag measurements from our CCFs (black markers), and our best-fit model (orange solid line). The {\sl bottom} panel displays the residuals of the fit, where residual=(data-model)/(measurement errors). The gray shaded region indicates the $1\sigma$ error region. The right vertical axis indicates the corresponding jet size scale for the time lags on the left most vertical axis and our best-fit parameters (using Equation~\ref{eq:z0}; units of $\times 10^{14}$ cm).
 }
 \end{figure}

\begin{equation}
\tau_{\rm lag}=
\frac{z_{{\rm norm}}\,(\frac{1}{{\nu_{\rm radio}}^\epsilon}-\frac{1}{{\nu_{\rm xray}}^\epsilon})\,(1-\beta\cos\theta)}{\beta\,c}\,.
\label{eq:lag}
\end{equation}
Here $\nu_{\rm radio}$ and $\nu_{\rm xray}$ (set to the middle of the X-ray band at 41 keV or $9.91\times10^{9}$ GHz, {although the fitting process is not very sensitive to the value of $\nu_{\rm xray}$}) represent the frequencies between which the time lags were measured. To normalize the jet size scale to radio frequency band relation, we 
can express $z_{\rm norm}$ in terms of 
the distance between the X-ray emitting region and the $\tau=1$ surface ($l_{\rm norm}$ in angular units of mas projected on the sky) at a specific radio frequency ($\nu_{\rm norm}$), yielding,
\begin{equation}
z_{\rm norm}=(1.49\times10^{13}\,{\rm cm})\, \frac{l_{\rm norm}\,D_{\rm kpc}}{\sin\theta}\,(\nu_{\rm norm})^\epsilon
\end{equation}
Based on the VLBI images of Cyg X-1 presented in \citealt{stir01} at $\nu_{\rm norm}=8.4$ GHz, we set a wide uniform prior on $l_{\rm norm}$ ranging from 0.01 mas to 15 mas (max distance to which jet emission was resolved in the VLBI images). 

\renewcommand\tabcolsep{12pt}
 \begin{table}
\caption{Best fit jet parameters}\quad
\centering
\begin{tabular}{ lc }
 \hline\hline
 {\bf Parameter}&{\bf Best-fit result} \\[0.15cm]
 \hline
  $\Gamma$&$2.59^{+0.79}_{-0.61}$\\[0.15cm]
  $\epsilon$&$0.40^{+0.05}_{-0.05}$\\[0.15cm]
  $l_{\rm norm}$ (mas)&$5.69^{+1.52}_{-1.62}$\\[0.15cm]
  $\beta^\dagger$&$0.92^{+0.03}_{-0.06}$\\[0.15cm]\hline
\end{tabular}\\
\begin{flushleft}
{$^\dagger$ $\beta$ represents the bulk jet speed and is not a fitted parameter (see \S\ref{sec:speed} for details). We instead fit for the bulk Lorentz factor, $\Gamma=(1-\beta^2)^{-1/2}$, and estimate the distribution of the corresponding bulk jet speeds by performing Monte Carlo simulations sampling from the posterior $\Gamma$ distribution 10000 times.}\\
\end{flushleft}
\label{table:bestfit}
\end{table}
\renewcommand\tabcolsep{6pt}

We fit Equation~\ref{eq:lag} to our measured X-ray/radio time lags using a Markov Chain Monte-Carlo (MCMC) algorithm \citep{for2013}. However, instead of fitting directly for $\beta$, we choose to fit for the bulk Lorentz factor, $\Gamma=(1-\beta^2)^{-1/2}$. This is because the MCMC algorithm performs better when there are not hard limits on model parameters (i.e., $\beta$ can only have values between 0 and 1). In our MCMC runs, we allow $\Gamma$, $\epsilon$, and $l_{\rm norm}$ to be free parameters, and sample from the known distance ($1.86\pm0.12$; \citealt{re11}) and inclination ($\theta=27.1\pm0.8$; \citealt{or11}) distributions. {We note that sampling from the {\em Gaia} DR2 distance distribution ($2.38^{+0.20}_{-0.17}$ kpc; \citealt{gand18}), rather than the radio parallax distance distribution, does not have a significant effect on the best-fit jet parameters (i.e., all parameters are consistent within the $1\sigma$ error bars for both fitting runs).}
The best-fit result is taken as the median of the one-dimensional posterior distributions and the uncertainties are reported as the range between the median and the 15th percentile (-), and the 85th percentile and the median (+), corresponding approximately to $1\sigma$ errors.
Our best-fit parameters are displayed in Table~\ref{table:bestfit} and Figure~\ref{fig:lag_fit}.

Prior to this study, the compact jet speed in BHXBs had never been directly measured (although \citealt{cas10} have placed direct lower limits on the jet speed, see below). While it is typically thought that compact jets are inherently less relativistic than the other form of jets detected in these systems, discrete jet ejecta (displaying bulk speeds as high as $\Gamma >2$, measured using proper motions derived through VLBI imaging; \citealt{hj00b,hjr95,fenhombel09}), past studies using indirect methods (e.g., scatter in radio/X-ray correlation, spread in accretion state transition luminosities\footnote{Note that the spread in the accretion state transition luminosities can only constrain the compact jet speed if the X-ray emission originates in the jet.}; \citealt{gallo2003,maca03,gleiss4,hm04}) to infer limits on the compact jet speed, have yielded at times conflicting results (e.g., bulk Lorentz factors of $\Gamma<2$ versus $\beta\Gamma>5$). 
Additionally, \citet{fender03} has shown that the inferred $\Gamma$ (measured from proper motions) depends strongly on the assumed distance to the source (which is not known accurately in most BHXBs), and \citet{millerj06} show that if the jets are not externally confined, the intrinsic Lorentz factors of BHXB jets could be as high as in AGN (up to $\Gamma\sim10$).
The new compact jet speed measurement derived from our time domain analysis presented here for Cyg X-1 shows a reasonably high bulk Lorentz factor of $\Gamma\sim 2.6$. A previous study on the compact jet in GX 339-4 also presented potential evidence for similarily high bulk Lorentz factors inferred from infrared/X-ray lags ($\Gamma>2$; \citealt{cas10}). Since the compact jet speed is likely to vary during an outburst \citep{vadawale2003,fenbelgal04,fenhombel09}, it is possible that our Cyg X-1 jet speed measurement samples the jet speed at the high end of the distribution in this source (rather than the mean). {In fact, \cite{fenray00} have shown evidence for varying radio-radio time lags in different observations of GRS 1915+105, which could be tracing a varying bulk Lorentz factor in the jet.} In this sense, jet speed constraints from different stages of an outburst in a single BHXB source are needed to understand the degree to which compact jet speed can change during an outburst. Similarly, jet speed constraints for multiple sources are needed to constrain the distribution of compact jet speeds across the BHXB population (as well as help determine if Cyg X-1 is an outlier in terms of higher compact jet speed).

Moreover, our time lag modelling also reveals that the size scale as a function of radio frequency in the Cyg X-1 jet seems to deviate from the expected $z\propto1/\nu$ relationship predicted by simple jet models \citep{blandford79}. In particular, we find that a shallower relationship ($\epsilon=0.40\pm0.05$, where $z\propto1/\nu^{\epsilon}$) is needed to explain our observed time lags.

Further, for a conical jet, the opening angle can be estimated based on the axial and transverse size scales of the jet according to,
\begin{equation}
\phi=\tan^{-1}\left({\frac{z_{\rm trans}}{z_{\rm axial}}}\right)
\label{eq:oa}
\end{equation}
Therefore, through applying estimates of $z_{\rm axial}$ (via our measured time-lags) and $z_{\rm trans}$ (via timescales derived from our Fourier analysis that may trace the transverse jet size scale), we can use Equation~\ref{eq:oa} to place constraints on the opening angle of the Cyg X-1 jet. Considering the size scales in the region of the jet sampled by the 11 GHz radio band, Equation~\ref{eq:z0} yields $z_{\rm axial}=3.1\times10^{14}$ cm, and we estimate $z_{\rm trans}=(0.1-1)\times10^{13}$ cm based on the smallest timescales over which we observe significant power in the PSD ($0.03$ Hz; see Figure~\ref{fig:psdr}) and the timescales at which we observe a significant drop in coherence in our 9/11 GHz cross-spectral analysis ($0.003$ Hz; see Figure~\ref{fig:csdr}). With these size scale estimates, we estimate the opening angle to be $\phi\sim0.4-1.8$ degrees. This constraint is consistent with both the VLBI upper limit ($<2$ degrees) and the work of \citet{heinz06}, the latter who present analytical expressions for jet parameters in Cyg X-1 (see Equation 9 of \citet{heinz06} for the opening angle expression).

Lastly, our modelling estimates that the radio jet emitting region (at 8.4 GHz) is located $l_{\rm norm}=5.69$ mas ($1.6\times10^{14}$ cm at 1.86 kpc) downstream from the X-ray emitting region. While the value of this parameter is currently unknown for Cyg X-1, it is of interest to compare our derived value to other estimates from different methods in the literature. Using Equation 4 of \citet{bro02}, we estimate that the radio photo-sphere at 8.4 GHz is located $4.4$ mas ($1.2\times10^{14}$ cm at 1.86 kpc) downstream, which is consistent with our value (within the $1\sigma$ errors). However, \citet{zd12} estimate the distance downstream to the 8.4 GHz emitting region is much smaller than our value at $\sim0.2$ mas ($\sim5.4\times10^{12}$ cm at 1.86 kpc). 
As the derived jet speed is highly dependent on the value of this downstream distance in our model, it is important to place improved constraints on this parameter to accurately measure jet speed with the method presented in this work.

We reiterate that all the results in this section hinge on the accuracy of our time lag measurements (see \S\ref{sec:ccf} for a discussion on the uncertainty in the time lags for the lower frequency radio bands) and the assumption that the bulk jet speed is constant with distance downstream. Therefore, the reader is cautioned that, while our results are intriguing, they come with caveats.

\section{Summary}
\label{sec:sum}

In this paper, we present the results of our simultaneous multi-band radio and X-ray observations of the BHXB Cyg X-1, taken with the VLA and {\em NuSTAR}. With these data, we extracted high time resolution light curves, probing timescales as short as seconds. The light curves display small amplitude flaring events, where the lower frequency radio emission appears to lag the higher frequency radio emission and any variability is consistent with being much more smoothed out at the lowest radio frequencies. This behaviour is consistent with emission from a compact jet, where higher radio frequencies probe closer to the black hole.

To better characterize the compact jet variability we observe in our light curves, and probe how this variability propagates down the jet, we performed timing analyses on our data, including Fourier domain analyses (PSDs and cross-spectral analyses between radio bands), as well as cross-correlation analyses. We summarize the key results of this analysis below.

Our radio PSDs show a monotonically decreasing trend with frequency, with no significant power at Fourier Frequencies $>0.03$ Hz ($< 30$ sec timescales). Over the Fourier frequencies that we sample, we do not observe a turnover in the radio PSDs, as was seen in recent studies analyzing infrared/optical PSDs of compact jet emission from the BHXB GX 339-4. However, upon comparing our radio PSDs to a past study probing longer timescales, we find that it is plausible that such a turnover in the PSDs could occur in the currently un-sampled $10^{-5}-10^{-3}$ Hz Fourier frequency range. As this turnover is thought to reflect the physical size scale of the emitting region in the jet, future studies with a set of longer ($>3$ hours) radio observations present the opportunity to detect such a turnover, and in turn use it to map out the jet size scale in different regions of the jet flow.

Our cross spectral analyses reveal that the higher frequency radio bands (9/11 GHz) are highly correlated over the same Fourier frequency range where we observe significant radio variability amplitude in our PSDs ($<0.01$ Hz), and a roughly constant time-lag is observed with Fourier frequency between these bands. These results are consistent with a propagation model whereby the initial radio signal is delayed and the variability at higher Fourier frequencies is smoothed out as the signal propagates down the jet. However, in the lower frequency radio bands (2.5/3.5 GHz) we observed very little correlation across all timescales. The loss of coherence could be a result of processes occurring on timescales shorter than the propagation time between emitting regions (such as turbulence in the jet flow) distorting the radio signals. 
We also discuss several other mechanisms that could cause a loss of coherence: a strong noise contribution in the radio light curves, other sources of emission contributing to our radio signals (e.g., interactions with the ISM, stellar wind absorption), and non-linear processes acting on the radio signals (e.g., shocks within the jet flow or at jet-ISM impact sites).

Our cross-correlation functions confirm the presence of a correlation between the radio and X-ray emission in Cyg X-1. We detect time lags of tens of minutes between the X-ray and radio bands, and use these measurements to solve for a jet speed of $\beta=0.92^{+0.03}_{-0.06}$ ($\Gamma=2.59^{+0.79}_{-0.61}$). Additionally, we also constrain how the jet size scale changes with frequency, finding a shallower relation than predicted by simple jet models ($z\propto1/\nu^{0.4}$), and estimate the jet opening angle to be $\phi\sim0.4-1.8$ degrees.

Overall, in this paper we have presented a detailed study of rapid compact jet variability (probing second to hour timescales) at radio frequencies in a BHXB. Our work here shows the power of time domain analysis in probing jet physics and displays the need for longer ($>3$ hours) continuous observations of BHXB jets across a range of electromagnetic frequency bands (especially including $>11$ GHz radio and sub-mm frequencies to bridge the gap to the infrared/optical bands). The combination of the techniques and tools developed in this study, as well as the improved capabilities of planned next generation instruments (such as the ngVLA and ALMA-2030), will make more of these radio time domain studies possible, not just in BHXBs, but in other jet-producing sources as well.

\section*{Acknowledgements}
The authors thank the anonymous referee for helpful feedback that improved the manuscript.
We also wish to thank Fiona Harrision for granting our {\em NuSTAR} DDT request, used to obtain the X-ray data analyzed in this paper. AJT thanks Julien Malzac for his helpful comments and suggestions on the work presented in this paper.
AJT is supported by an Natural Sciences and Engineering Research Council of Canada (NSERC) Post-Graduate Doctoral Scholarship (PGSD2-490318-2016). AJT, BET, and GRS are supported by NSERC Discovery Grants. JCAMJ is the recipient of an Australian Research Council Future Fellowship (FT140101082). PG is supported by STFC (ST/R000506/1). GRS, PC, and PG acknowledge the support of the International Space Science Institute -- Beijing (ISSI-BJ) for team meetings on ``Understanding multi-wavelength rapid variability: accretion and jet ejection in compact objects''. The authors also thank the Lorentz Center in Leiden for hosting the workshop ``Paving the way to simultaneous multi-wavelength astronomy'', where this project was first developed.
The authors acknowledge the use of Cybera Rapid Access Cloud Computing Resources and Compute Canada WestGrid Cloud Services for this work. We also acknowledge support from the SKA/AWS AstroCompute in the Cloud Program, whose resources were used to create and test the {\sc casa} variability measurement scripts used in this work. The National Radio Astronomy Observatory is a facility of the National Science Foundation operated under cooperative agreement by Associated Universities, Inc. This work made use of {\em NuSTAR} mission data, a project led by the California Institute of Technology, managed by the Jet Propulsion Laboratory, and funded by NASA. We acknowledge extensive use of the {\sc stingray} python package (\url{https://stingray.readthedocs.io/en/latest/}), as well as the ZDCF codes of \cite{alex97}, in our analysis.




\bibliography{ABrefList} 



\appendix

\section{Radio Calibrator Light Curves}
\label{sec:appendix}
\begin{figure*}
\begin{center}
 \includegraphics[width=1\textwidth]{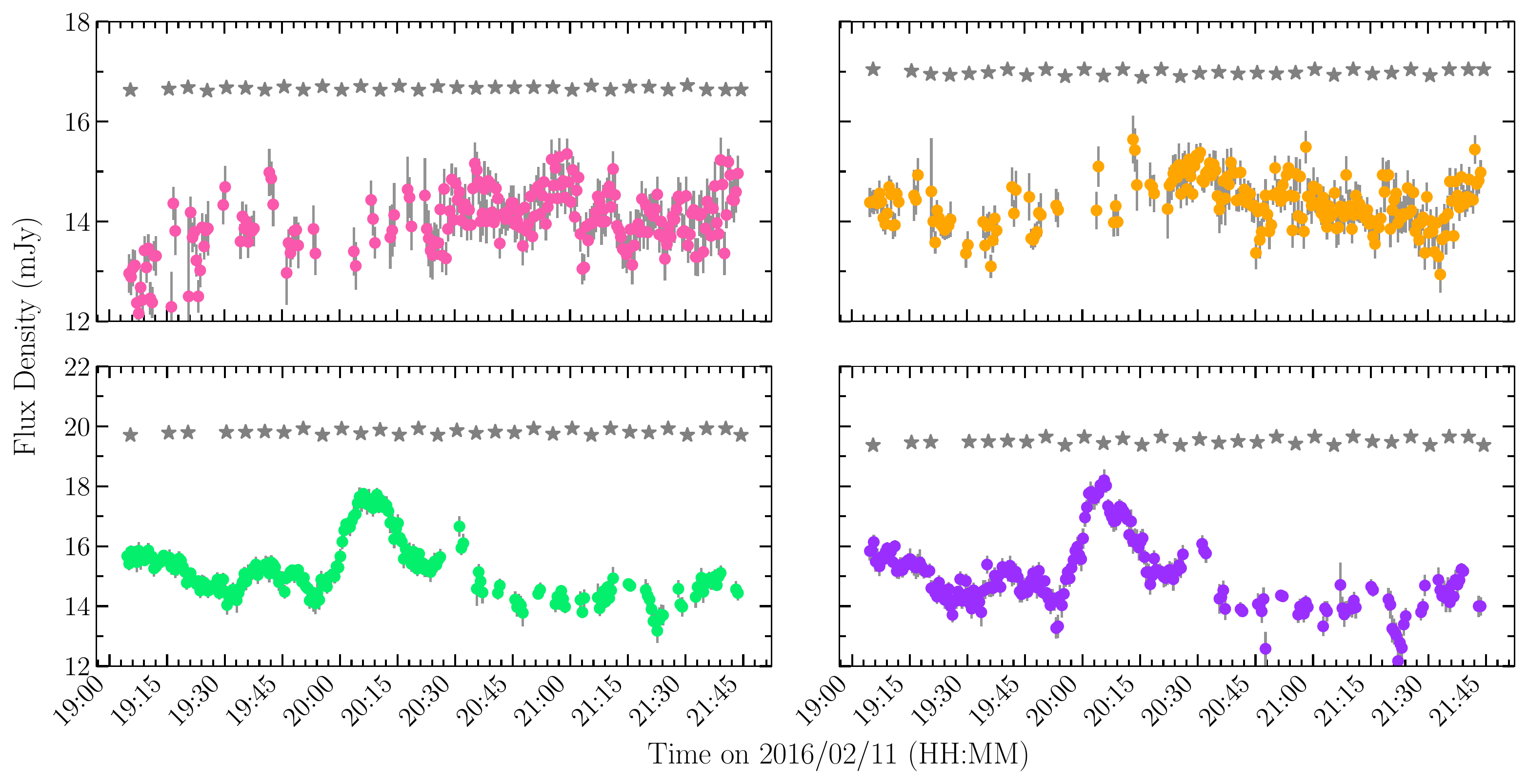}
 \caption{\label{fig:calslc} 
Multi-frequency light curves of Cyg X-1 and our calibrator source (J2015$+$3710); 2.5\,GHz ({\sl top left}), 3.5\,GHz ({\sl top right}), 9\,GHz ({\sl bottom left}), and 11\,GHz ({\sl bottom right}) bands. In all panels, the calibrators are plotted as gray star symbols and the total flux densities of the calibrators are scaled down for clarity in the plots (the average flux densities of J2015$+$3710 are 2916.4, 3481.5, 3963.3, and 3803.9 mJy in the 2.5, 3.5, 9, and 11 GHz bands). The calibrator observations in all bands display a relatively constant flux density throughout our observations, indicating that the variations we observe in Cyg X-1 are most likely intrinsic to the source, rather than due to atmospheric or instrumental effects.}
\end{center}
\end{figure*}

Given the flux variations we detected in our VLA radio frequency data of Cyg X-1, we wished to check the flux calibration accuracy of all of our observations on short timescales and ensure that the variations observed in Cyg X-1 represent intrinsic source variations rather than atmospheric or instrumental effects. Therefore, we created time resolved light curves of our calibrator source, J2015+3710 (see Figure~\ref{fig:calslc}).

We find that J2015+3710 displays a relatively constant flux density throughout our observations in all the sampled bands, with any variations ($<$1\% of the average flux density at all bands) being a very small fraction of the variations we see in Cyg X-1.
Based on these results, we are confident that our light curves of Cyg X-1 are an accurate representation of the rapidly changing intrinsic flux density of the source.

\section{X-ray Light curves}
\label{sec:xraylc_ex}

\begin{figure}
   \centering
   { \includegraphics[width=1\columnwidth]{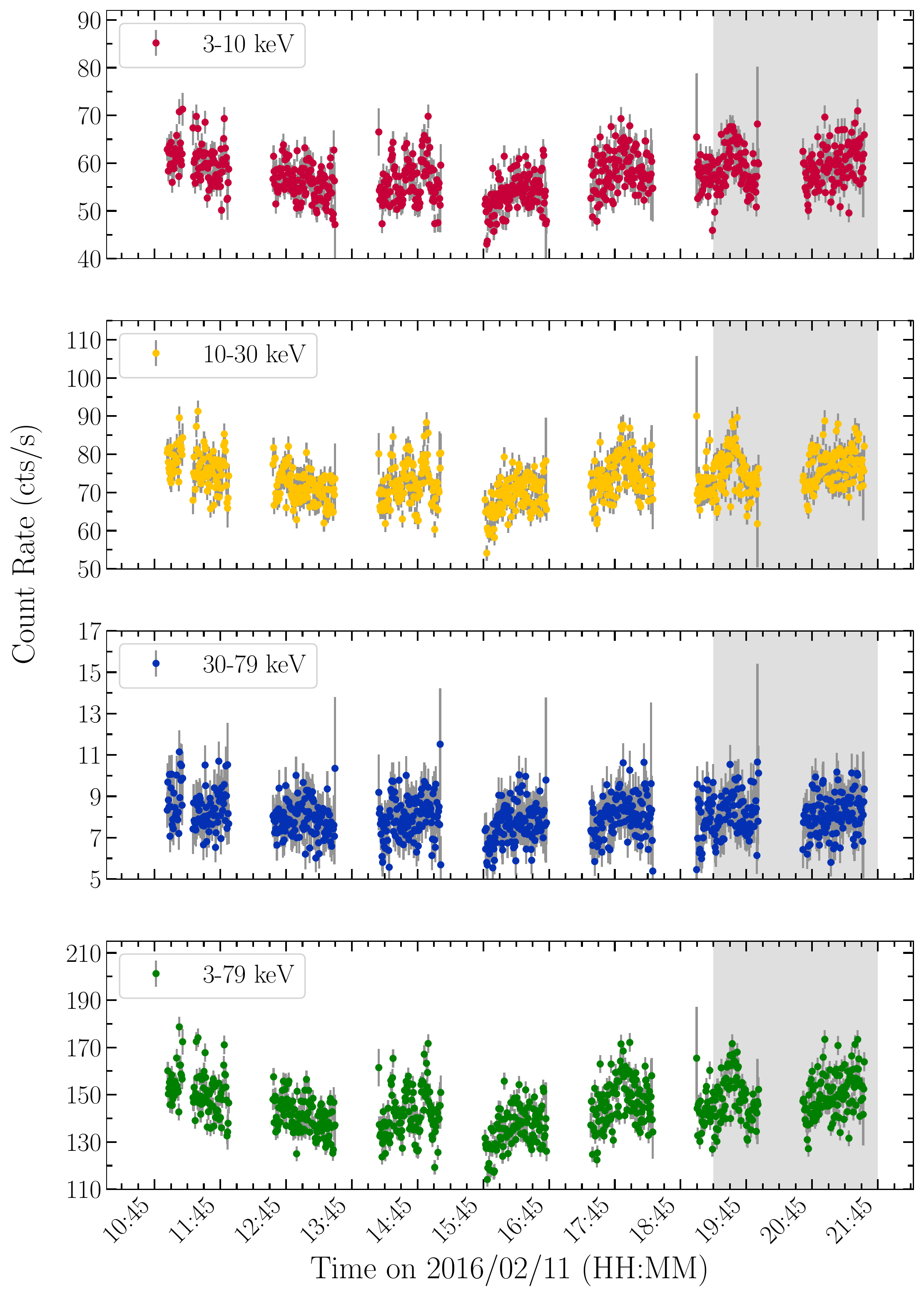}} \\
 \caption{ \label{fig:Xlc} Extended version of Figure~\ref{fig:cyglc}, where our complete {\em NuSTAR} X-ray  light curves of the BHXB Cyg X-1 are shown (with 30 sec time-bins) in all bands, including the full 3--79 keV band. The gray shading represents the period of overlap between the VLA and {\em NuSTAR} observations, in which simultaneous X-ray and radio data were obtained.
 }
 \end{figure}

In this section we display an extended version of our {\em NuSTAR} X-ray light curves covering the full observation period, including the period prior to the overlap with our VLA radio observations (see Figure~\ref{fig:Xlc}).

\section{PSD white noise subtraction}
\label{sec:app2}
\begin{figure*}
   \centering
   
   \subfloat{ \includegraphics[width=0.9\columnwidth,height=5cm]{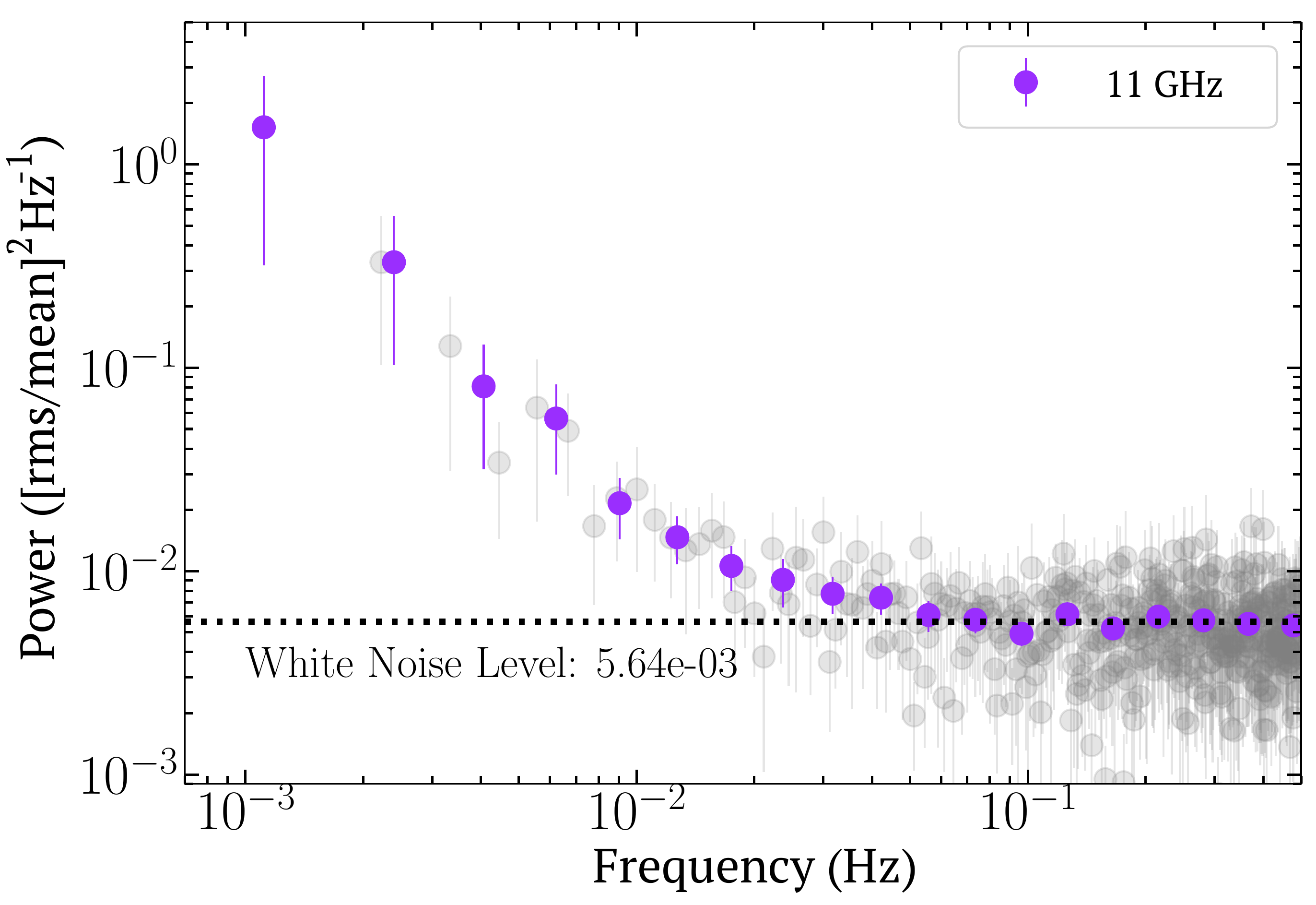}}\quad\quad
   \subfloat{ \includegraphics[width=0.9\columnwidth,height=5cm]{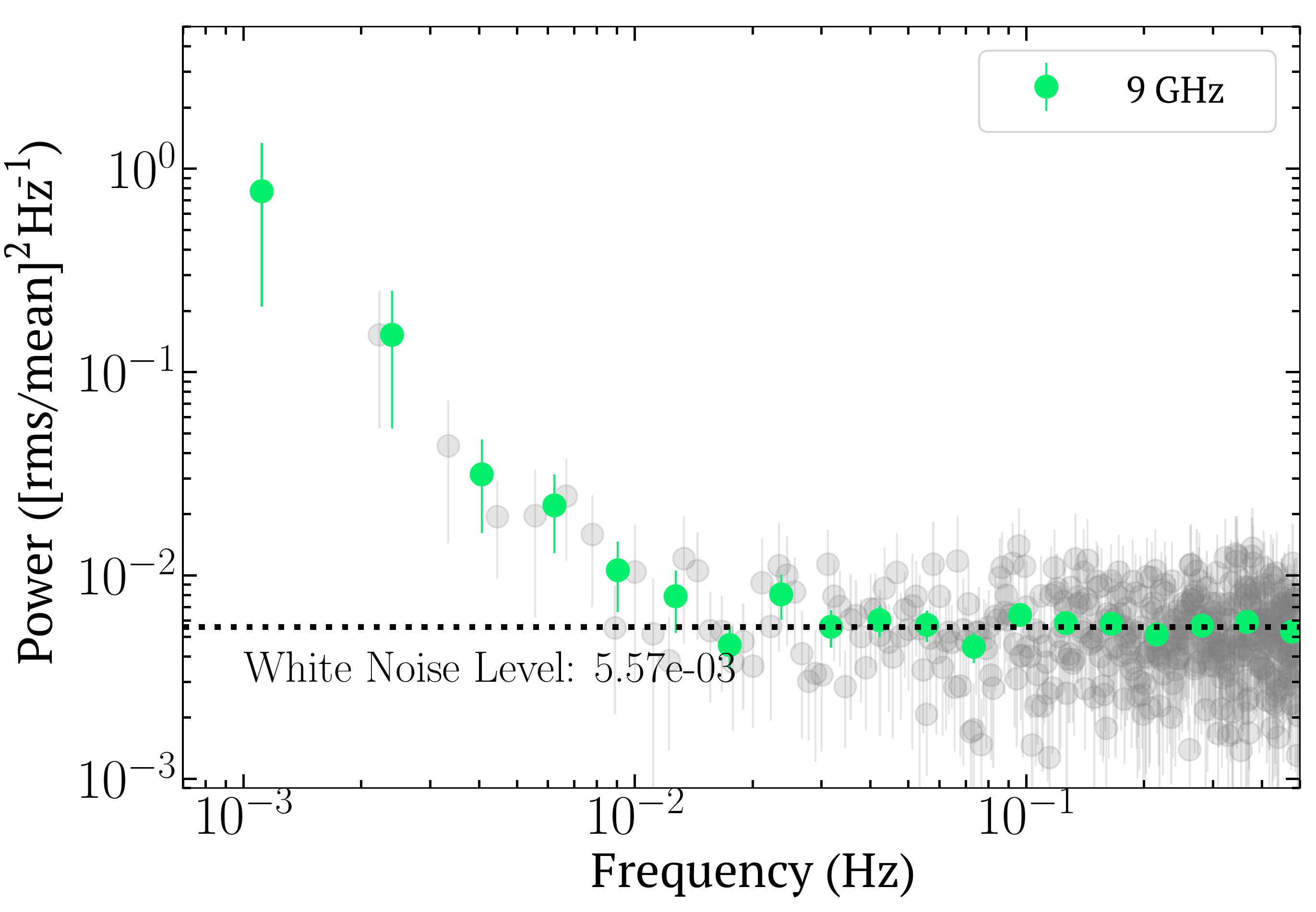}}\\ \vspace{0.4cm}
   \subfloat{ \includegraphics[width=0.915\columnwidth,height=5cm]{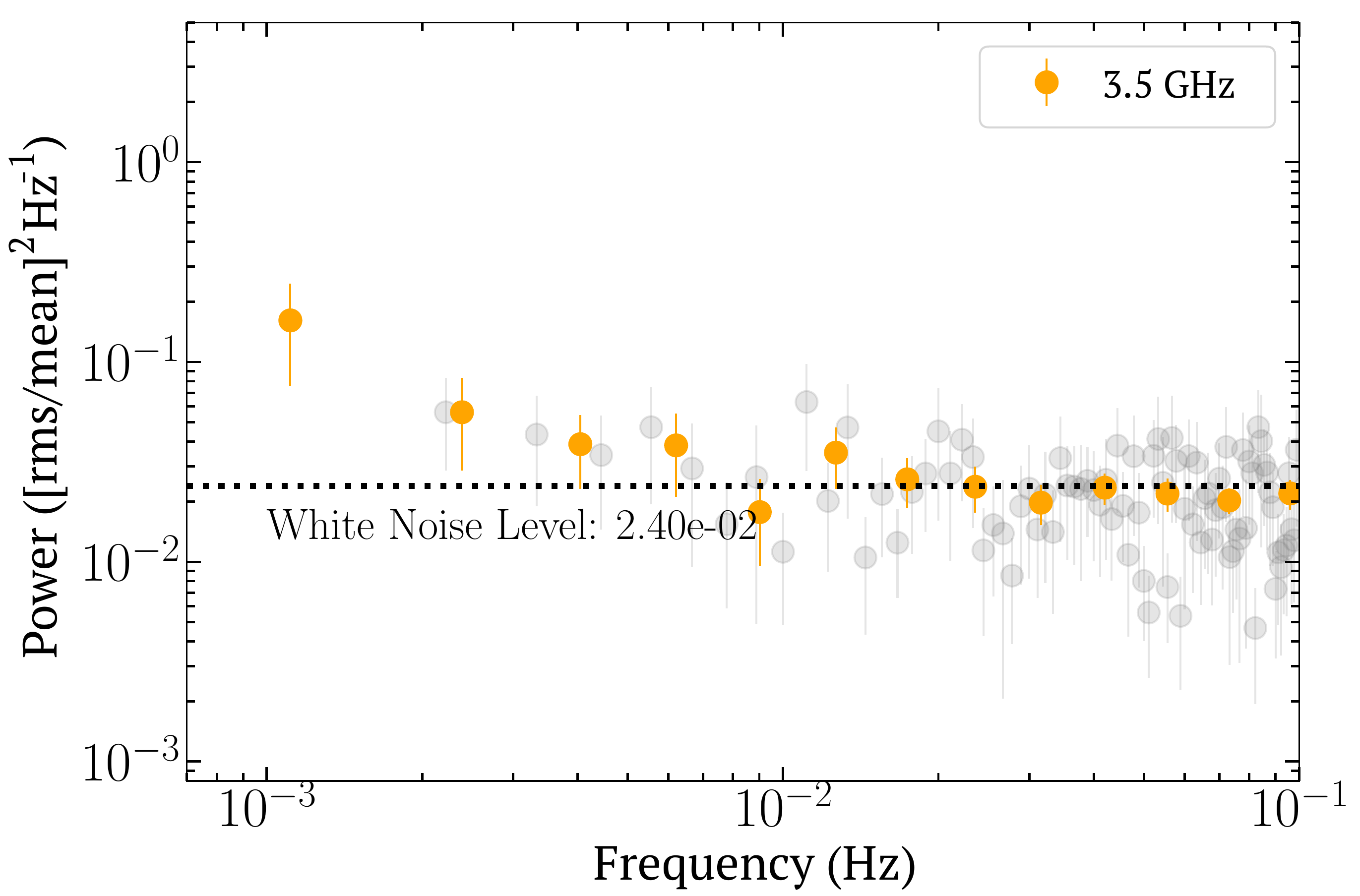}}\quad\quad
   \subfloat{ \includegraphics[width=0.915\columnwidth,height=5cm]{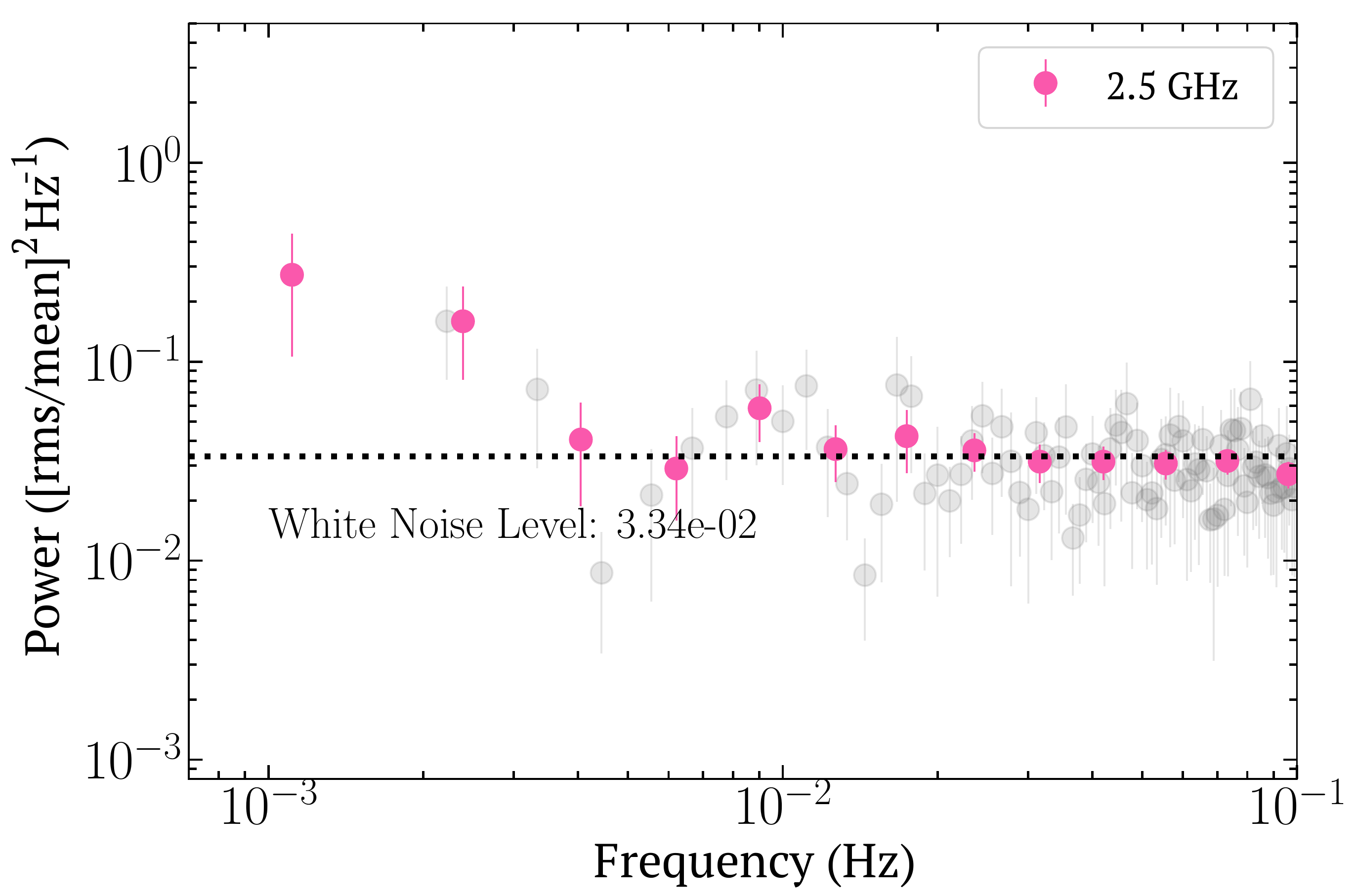}} 
 \caption{\label{fig:psdr_wn} Radio PSDs of Cyg X-1 prior to white noise subtraction. In each panel, the gray-scale markers represent the PSDs before re-binning in frequency, and the coloured markers represent the PSDs after re-binning. The white noise level (in units of power, displayed on the vertical axis) is also indicated in each panel. The PSDs in the 9/11 GHz bands are created from data imaged with 1 sec time-bins, while the 2.5/3.5 GHz PSDs are created from data imaged with 5 sec time-bins.
 }
 \end{figure*}

In this section we display the radio PSDs prior to white noise subtraction, and indicate the measured white noise levels (see Figure~\ref{fig:psdr_wn}). To determine whether the white noise could be intrinsic to the source (and in turn if subtracting the white noise in our PSDs is a valid practice in this case), we compared the white noise levels shown in Figure~\ref{fig:psdr_wn} to the rms noise levels in the individual time-bin images that make up each light curve. We find that the white noise levels closely match the average rms noise levels in the images at each radio frequency band. This indicates that the source of the white noise in the PSDs is likely not intrinsic to the source but rather due to atmospheric/instrumental effects (which govern the rms noise levels in radio frequency images).

%


\bsp	
\label{lastpage}
\end{document}